\documentclass[structabstract]{aa}  
\usepackage{graphicx}
\usepackage{lscape}
\usepackage{longtable}
\usepackage{natbib}
\usepackage{color}
\usepackage{array} 
\bibpunct{(}{)}{;}{a}{}{,} 
\usepackage{txfonts}
\begin{document}
%
	\title{Kepler-91b: a planet at the end of its life}
	\subtitle{Planet and giant host star properties via light-curve variations  \thanks{Based on observations collected at the German-Spanish Astronomical Center, Calar Alto, jointly operated by the Max-Planck-Institut fur Astronomie (Heidelberg) and the Instituto de Astrof\'isica de Andaluc\'ia (IAA-CSIC, Granada).}}

  \author{J. Lillo-Box
          \inst{1}
          , D. Barrado\inst{1,2},  A. Moya\inst{1}, B. Montesinos\inst{1}, J. Montalb\'an\inst{3}, A. Bayo\inst{4,5}, M. Barbieri\inst{6}, C. R\'egulo\inst{7,8},  L. Mancini \inst{4}, H. Bouy\inst{1}, and T. Henning\inst{4}  }

  \institute{Departamento de Astrof\'isica, Centro de Astrobiolog\'ia (CSIC-INTA), ESAC Campus 28691 Villanueva de la Ca\~nada (Madrid), Spain\\  \email{Jorge.Lillo@cab.inta-csic.es}        \and
Centro Astron\'omico Hispano-Alem\'an (CAHA). Calar Alto Observatory, c/ Jes\'us Durb\'an Rem\'on 2-2, 04004, Almer\'ia, Spain.               \and
Institut d'Astrophysique et G\'eophysique de l'Universit\'e de Li\`ege, All\'ee du six Ao\^ut, 17 B-4000 Li\`ege, Belgium \and
Max Planck Institute for Astronomy, K\"onigstuhl 17, 69117 Heidelberg, Germany  \and 
European Southern Observatory, Alonso de C\'ordova 3107, Vitacura, Santiago, Chile \and
Osservatorio Astronomico di Padova. Vicolo Osservatorio 5 - 35122 - Padova, Italy.  \and
Instituto de Astrof\'isica de Canarias, 38205 La Laguna, Tenerife, Spain \and
Dpto. de Astrof\'isica, Universidad de La Laguna, 38206 La Laguna, Tenerife, Spain \\
      }


  \date{Accepted for publication in A\&A on December 5th, 2013}
  
\titlerunning{Kepler-91b: a giant planet at the end of its life}
\authorrunning{Lillo-Box et al.}


  \abstract
  {The evolution of planetary systems is intimately linked to the evolution of their host star. Our understanding of the whole planetary evolution process is based on the large planet diversity observed so far. To date, only few tens of planets have been discovered orbiting stars ascending the Red Giant Branch.  Although several theories have been proposed, the question of how planets die remains open due to the small number statistics, making clear the need of enlarging the sample of planets around post-main sequence stars.}
  {In this work we study the giant star Kepler-91 (KIC 8219268) in order to determine the nature of a transiting companion. This system was detected by the \textit{Kepler} Space Telescope, which identified small dims in its light curve with a period of $6.246580\pm0.000082$ days. However, its planetary confirmation is needed due to the large pixel size of the \textit{Kepler} camera which can hide other stellar configurations able to mimic planet-like transit events.} 
  { We analyse \textit{Kepler} photometry to: 1) re-calculate transit parameters, 2) study the light-curve modulations, and 3) to perform an asteroseismic analysis (accurate stellar parameter determination) by identifying solar-like oscillations on the periodogram. We also used a high-resolution and high signal-to-noise ratio spectrum obtained with the Calar Alto Fiber-fed \'Echelle spectrograph (CAFE) to measure stellar properties. Additionally, false-positive scenarios were rejected by obtaining high-resolution images with the AstraLux lucky-imaging camera on the 2.2 m telescope at the Calar Alto Observatory.}
  { We confirm the planetary nature of the object transiting the star Kepler-91 by deriving a mass of $ M_p=0.88^{+0.17}_{-0.33} ~M_{\rm Jup}$ and a planetary radius of  $R_p=1.384^{+0.011}_{-0.054}  ~R_{\rm Jup}$. Asteroseismic analysis produces a stellar radius of $R_{\star}=6.30\pm 0.16   ~R_{\odot}$ and a mass of $M_{\star}=1.31\pm 0.10 ~  M_{\odot} $. We find that its eccentric orbit ($e=0.066^{+0.013}_{-0.017}$) is just $1.32^{+0.07}_{-0.22} ~ R_{\star}$ away from the stellar atmosphere at the pericenter. We also detected three small dims in the phase-folded light-curve. The combination of two of them agrees with the theoretical characteristics expected for secondary eclipse.}
  {Kepler-91b could be the previous stage of the planet engulfment, recently detected for BD+48 740. Our estimations show that Kepler-91b will be swallowed by its host star in less than 55 Myr. Among the confirmed planets around giant stars, this is the planetary-mass body closest to its host star.  At pericenter passage, the star subtends an angle of $48^{\circ}$, covering around 10\% of the sky as seen from the planet. The planetary atmosphere seems to be inflated probably due to the high stellar irradiation.}
  


  \maketitle
%


\section{Introduction}

From a theoretical point of view, giant planets around red giant stars have been extensively studied in recent years \citep{burkert07,villaver09,kunitomo12,passy12}. Observationally, few tens of exoplanets have been found so far orbiting these evolved stars \citep[e.g, ][]{johnson07,adamow12,jones13}. In particular, the discovery of planets around K and G giants is crucial { for} planet formation theories. These stars { evolved from F- and A-type Main-Sequence stars, for which} accurate radial velocity studies are difficult (due to the small number of absorption lines present in their spectrum) and { therefore}, confirmation of planet candidates becomes hard.  { Since, as a result of this,} very few planets have been found orbiting F and A parent stars, K and G giants (with deeper absorption lines) can help {to better constrain the demography of planets around early-type stars.}

{ In addition}, there is a paucity of planets with short periods around stars ascending the Red Giant Branch \citep[RGB,][]{johnson07}. This desert has been theoretically studied by \cite{villaver09}, who concluded that it can be explained by planet disruption/engulfment during the ascent along the RGB (although other mechanisms are possible). However, as they state, these results are based on a limited sample of confirmed exoplanets around RGB stars.  Therefore the detection of extremely close-in planets around {post Main-Sequence (giants)} stars can then constrain theoretical models about how planets are destroyed by their hosts.

{ In this context}, we present the confirmation of the planetary nature of the {\it Kepler} Object of Interest KOI-2133{b} (KIC 8219268b and hereafter Kepler-91b), { a close-in planet orbiting a K3 star in the giant branch.} We achieve this confirmation by exploiting the high-precision photometry provided by the {\it Kepler} mission \citep{borucki10} { and complementary data}. { The accuracy of the {\it Kepler} light-curve} allows us to detect small variations (of the order of tens of parts per million) in the out-of transit signal of the host star. {Whenever a companion is present,} the photometric modulation is known to be caused by the combination of three main factors: reflected/emitted light from the planet, ellipsoidal variations (or tidal distortions) induced by the planet on the star, and Doppler beaming due to the reflex motion of the star induced by the presence of a massive companion. These effects have been recently used to confirm a handful of transiting planets such as KOI-13 \citep{shporer11,mazeh12,mislis12}, HAT-P-7b \citep{borucki09,welsh10}, and {Kepler}-41b \citep{quintana13}. It is only possible to detect close-in giant planets with this method due to the subtle induced modulations. {Note that other physical processes such as stellar activity or pulsation can also modulate the stellar light-curve.} Besides, even if the modulation is indeed produced by a companion, this technique does not provide the absolute value for its mass but instead the companion-to-host mass ratio. Thus, it is crucial to obtain the most accurate parameters for the host star. This is achieved (when possible) using the asteroseismology \citep[see, for instance, ][]{aerts10,mathur12}. The launch of space-borne high-accuracy photometers such CoRoT \citep{baglin06} and {\it Kepler} has permitted to obtain long-term very accurate photometry for several hundred thousand stars, which has implied the rapid growth of this discipline.

A recent paper by \cite{esteves13} discards Kepler-91b as a planet candidate due to their finding of a large albedo corresponding to a self-luminous object. In this paper we perform careful analysis of all the public data and our own observations, and we firmly conclude that the transiting object is actually a very close-in hot-Jupiter planet in a stage previous to be engulfed by its host star. 

The paper is organized as follows. In section \S~2, we explain all the observational data available for this object, including our high-resolution images (\S~2.1) and our high-resolution and high signal-to-noise ratio spectrum (\S~2.2). We perform an exhaustive analysis of the host star properties in section \S~3, comprising spectral energy distribution (\S~3.3) high-resolution spectrum (\S~3.4), and an asteroseismic study of the object (\S~3.5). We then analyse the signals induced by the planet candidate into the stellar light-curve in section \S~4, including a new primary transit fit (\S~4.1) and a detailed fitting of the light-curve modulations (\S~4.2). Other aspects of the light-curve such as the possible secondary eclipses are discussed in section \S~4.3 and the final discussion and conclusions of the paper are presented in \S~5. { Throught out this paper we will refer to the host star as Kepler-91, and add ``b" when talking about the planet candidate. }


\section{Observations and data reduction}

\subsection{High spatial resolution: lucky imaging with CAHA/AstraLux}

We applied the lucky imaging technique to the planet host candidate
Kepler-91 in order { to search for a possible stellar companion by achieving }diffraction-limited resolution images. Due to the large pixel size of the {\it Kepler} camera (3.98 arcsec/pixel) and the much larger aperture (6-10 arcsec), high-resolution imaging is crucial to discard other possible stellar configurations mimicking the planetary transit \citep[see, for example,][]{daemgen09,lillo-box12,adams13}.

We used the AstraLux North instrument mounted on the 2.2m telescope at the Calar Alto Observatory (CAHA, Almer\'ia, Spain). The observations were performed on May 25th, 2012, {with a mean seeing of 0.8 arcsec}.  We obtained 30000 images of 50 milliseconds exposure time in the full CCD array of the camera ($24\times24$ arcsec$^2$). Data cube images were reduced using the online pipeline of the instrument \citep{hormuth07}, which performs basic { cosmetic and preparatory tasks}, selects the highest quality images, combines the best 1.0\%, 2.5\%, 5.0\%, and 10\% frames with the highest Strehl ratios \citep{strehl1902}, calculates the shifts between the single frames, performs the stacking, and re-samples the final image to half the pixel size (i.e., from 0.0466''/pixel to 0.0233''/pixel).

\begin{figure}[h]
   \centering
   \includegraphics[width=0.5\textwidth]{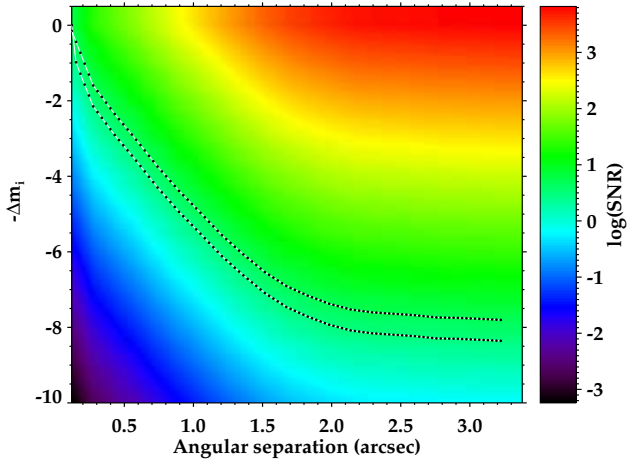} 
      \caption{{ Contrast map} of our high-resolution image taken with the 2.2m telescope plus AstraLux at Calar Alto Observatory with the 10\% of selection rate. { Colour code corresponds to the SNR with which we would detect a theoretical source with a magnitude difference $\Delta m_i$ (y-axis) at the corresponding angular separation (x-axis)}. The two dotted lines represent the $3\sigma$ (lower line) and $5\sigma$ (upper line) contours.}
         \label{sens}
   \end{figure}

In Fig.~\ref{sens} we show the sensitivity {map} for the 10\% selection rate image. According to our experience with the instrument, this rate provides the greatest image quality (regarding the interplay between magnitude difference and angular resolution). We refer the reader to \cite{lillo-box12} for a detailed explanation about the determination of the signal-to-noise ratio for each pair of angular separation and magnitude difference. Table~\ref{tab:astralux} shows the sensitivity limits for the four selection rate images within 1.5 arcsec angular separation from the target star. 

\subsection{High spectral resolution: \'echelle data with CAHA/CAFE}

We have obtained a high-resolution, high signal-to-noise ratio (SNR) spectrum of Kepler-91 by using the Calar Alto Fiber-feb \'Echelle spectrograph \citep[CAFE, ][]{aceituno13} on the 2.2m telescope. This instrument consists of a high dispersion spectrograph ($R=62000$) located in an isolated, controlled chamber and fed with a 2.0 arcsec diameter fiber. The stability of the instrument has been proven in \cite{aceituno13} where the authors reproduce the expected radial velocity curve of the planet TrEs-3b ($V=12.5$\,mag) with the data collected during commissioning.

\begin{figure}[h]
   \centering
   \includegraphics[width=0.5\textwidth]{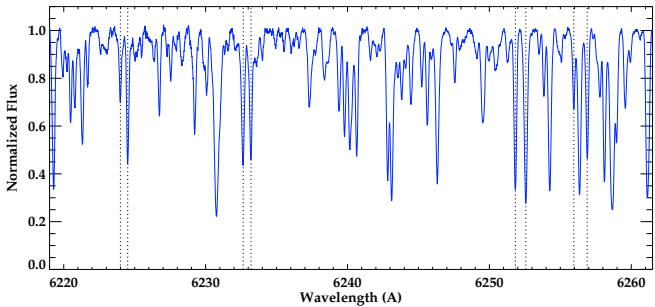}
      \caption{Small range of the high-resolution, high signal-to-noise ($SNR=123$ at 5800 \AA ) spectrum of Kepler-91 obtained with the CAFE \'echelle spectrograph at Calar Alto Observatory. We show the part of the spectrum used in the determination of the stellar metallicity (see section \S~\ref{sec:metallicity}) and mark with dotted lines the doublets used to estimate the effective temperature (see section \S~\ref{sec:teff}). The spectrum has been shifted to the rest frame by correcting for the barycentric velocity.}
         \label{fig:spectrum}
   \end{figure}

The { data were} reduced using the improved pipeline\footnote{See Appendix on http://www.caha.es/CAHA/Instruments/CAFE/\\ /cafe/CAFE.pdf} provided by the observatory which delivers a fully reduced spectrum \citep[see details in ][]{aceituno13}. { A small range of the spectrum is shown in Fig.~\ref{fig:spectrum}.}

\subsection{{\it Kepler} photometry: data handling}

The {\it Kepler} telescope has been almost continuously collecting data from the same field of view between March 2009 and April 2013. Its individual exposure { time is} 6.02 s with a 0.52 s readout time \citep{gilliland10b}. For long-cadence targets (which is our case), the telescope integrates over 270 exposures resulting in a total time resolution of 29.4 minutes per data-point. Time series are publicly available through the {\it Kepler} MAST (Mikulski Archive for Space Telescopes) webpage\footnote{http://archive.stsci.edu/kepler/}.

{ Taking advantage of} the large number of photometric points (around 52000), we decided not to remove possible outliers or de-trend the Pre-search Data Conditioning Simple Aperture Photometry flux \citep[PDCSAP,][]{smith12,stumpe12} { given the unknown} nature of these possible trends. { However, in} order to compare how de-trending { could} improve the quality of our data, we applied an iterative rejection process. First, the entire data-set { has been} split into continuous sections (i.e., regions without temporal gaps). Each section { was then} fitted with a fifth degree polynomial. Typical duration of the different sections are around 25-30 days (roughly one third of one {\it Kepler} quarter). Then, we { divided} our data by this fitted model and { removed} data points above $3\sigma$. We { iterated this process} until no { further} outliers { were} detected. While the standard deviation of the {raw PDCSAP} flux ($\approx 1152$ days) is $\sigma_{raw}=400$ ppm, the { resulting} cleaned light curve yields $\sigma_{corrected}=380$ ppm. Since the improvement is below 5 \%, we preferred not to apply any correction to the PDCSAP flux to { prevent} possible artificially-added trends.


\section{Properties of the host star  \label{sec:HostProperties}}

In the characterization of exoplanet properties, it is crucial to obtain the most accurate host-star parameters (radius, mass, effective temperature, age, etc.). { The inference of} both orbital and physical properties of the planet strongly depends on how well the stellar parameters are known \citep{seager03}. We have used our { wealth of data on} Kepler-91 to accurately determine these physical parameters following independent methods: model fit to the spectral energy distribution (SED), model fitting the high signal-to-noise spectrum, individual characterization of particular spectral lines, asteroseismology { and comparison with} isochrones and evolutionary tracks. Table~\ref{table:hostpars} provides a summary of all stellar parameters derived by these methods. Descriptions of each of them follows.

\subsection{Ancillary data and previous parameter estimations \label{sec:AncillaryPars}}
The stellar parameters of  Kepler-91 have been previously estimated by several methods that yield quite different results as shown in Table~\ref{table:hostpars} and summarized hereafter.

The {\it Kepler} Input Catalog (KIC, {\it Kepler} Team, 2009) provided photometric parameters for the whole sample of KOIs based on $u$, $g$, $r$, $i$, $z$, $J$, $H$, and $Ks$ magnitudes obtained by \cite{latham05}. { Their estimations are}: $T_{\rm eff}=4712$~K, $\log g=2.852$ [cgs], $\rm{[Fe/H]}=0.509$, and $E(B-V)=0.137$~mag or $A_V=0.425$~mag for $R_V=3.1$ (no errors are provided in this catalog). 

\cite{pinsonneault12} presented effective temperature corrections for the {\it Kepler} targets using SDSS colours and reported a value of $T_{\rm eff}=4837\pm96$ K for Kepler-91 ( assuming a metallicity of $\rm{[Fe/H]}= -0.2$). Surface gravity corrections were applied in that work to account for the evolved state of this target. We note that within the small sample of giants with spectroscopic information in the {\it Kepler} catalogue, the discrepancies between the SDSS temperature and the spectroscopic temperature range between -100 K and +400 K.  

In \cite{batalha13}, KIC effective temperature and $\log g$ are used as initial values for a parameter search using the Yonsei-Yale stellar evolution models, yielding { refined} values for the stellar mass and radius. { For the latter one, the authors reported} $R_{\star}=9.30~R_{\odot}$, which, together with the surface gravity, yields $M_{\star}=2.25~M_{\odot}$ (no errors provided), { implying} a mean stellar density of $\rho_{\star}=3.9~\rm{kg/m^3}$.  

   \begin{figure*}[ht]
   \centering
   \includegraphics[width=0.7\textwidth, angle=-90]{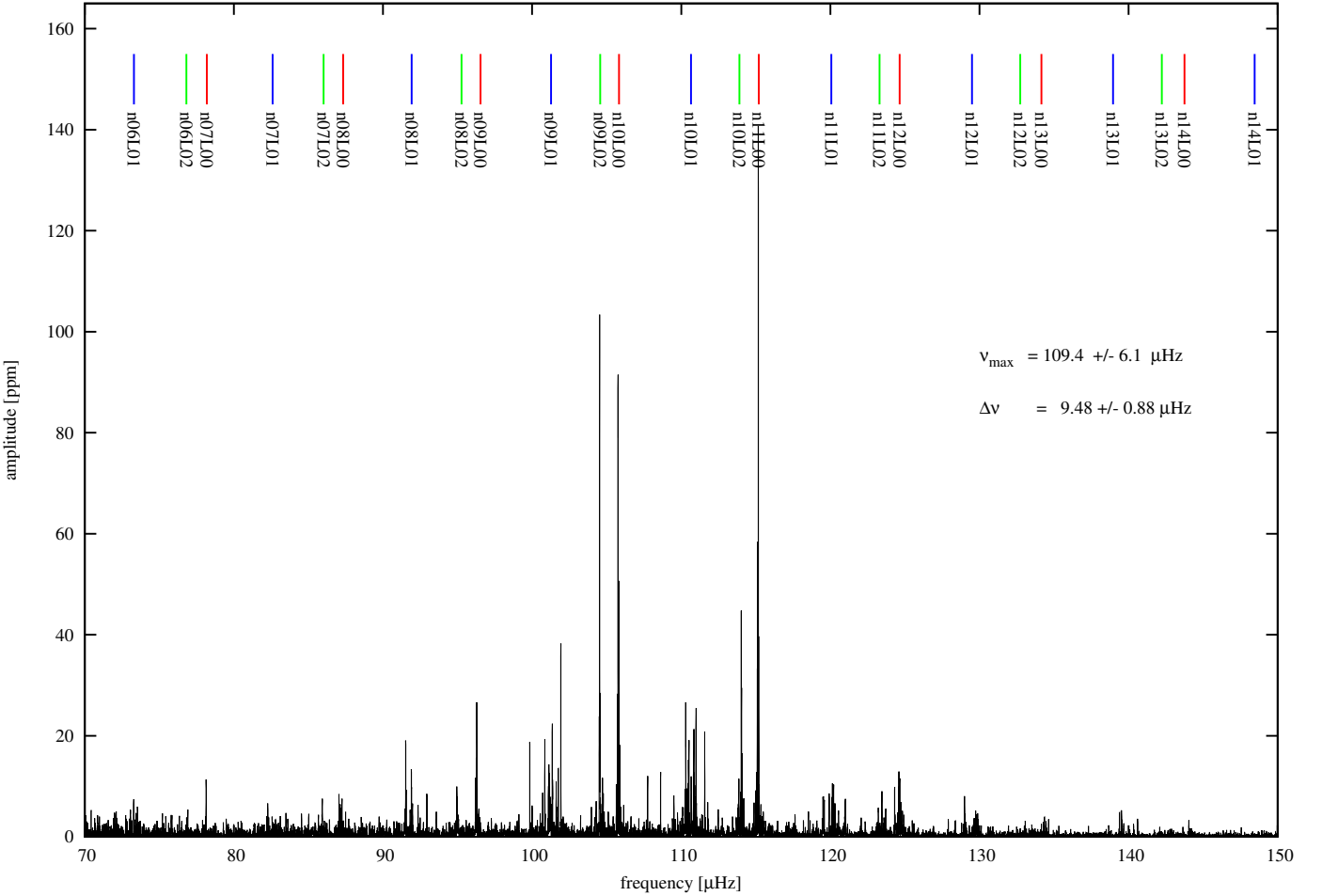}
      \caption{Power spectrum of the light curve, in a region centred on the maximum of the oscillations. The upper part of the figure shows the various modes identified for increasing value of $n$: in red the modes with $l=0$, green $l=2$, blue $l=1$. The black dotted line represents the heavy smoothed power spectrum. }
         \label{fig:mauro}
   \end{figure*}

Finally, since the power spectrum of the {\it Kepler} light curve of Kepler-91 presents the typical set of frequency peaks generated by giant stars (see Fig.~\ref{fig:mauro}), an asteroseismic analysis was performed by \cite{huber13}. Their results provided a mean density  of $\rho_{\star}=6.81\pm 0.32$~kg/m$^{3}$. Together with  the  spectroscopically derived effective temperature ($T_{\rm eff}=4605\pm97$~K) and asteroseismic relations \citep[i.e., ][see also section \S~3.5.1]{kjeldsen95}, they derived $R_{\star}=6.528\pm0.352~R_{\odot}$,  $M_{\star}=1.344\pm0.169~M_{\odot}$ which translates into a surface gravity of $\log g=2.94\pm0.17$.  By analysing three high-resolution spectra with the Stellar Parameter Classification technique \citep{buchhave12} the authors also estimated a $v\sin{i}=3.2\pm0.5$~km~s$^{-1}$, and a metallicity of $\rm{[Fe/H]} =0.29\pm0.16$. However, we must take into account that these asteroseismic relations (called scaling relations, see details in section \S~\ref{sec:asteroseismology}) have been obtained by comparing mainly stars with solar abundances, and the impact of a very different internal metallicity is not fully understood. Owing to the over-solar metallicity obtained by \cite{huber13}, a more detailed work on the fit of the oscillating frequencies is needed to obtain more accurate values of these parameters. In section \S~\ref{sec:IndividualModelling}, we perform an individual frequency modelling to obtain the most accurate stellar parameters possible with the current data.

We have used a complete set of available data for this object (photometry, {\it Kepler} light curve, and our own high-SNR spectrum) to compute \textit{self-consistent} values for the stellar and orbital parameters. In sections \S~3.2-3.5 we give details of these determinations.

\subsection{Multiplicity and projection effects study based on high spatial resolution observations}

{ We used our AstraLux high-resolution image to} calculate the chance-aligned probability of a non-resolved eclipsing binary in our high-resolution image { as a function of} angular separation ($\alpha$) and magnitude depth ($\Delta m$).  { We determined} the density of stars $\varrho$ at a given galactic latitude $b$ and magnitude difference with the target, $\varrho=\varrho(b, m_i, m_i+\Delta m)$. The number of possible chance-aligned sources within this angular separation is thus $N=\pi \alpha^2 \varrho$. We have calculated the density $\varrho$ by following the scheme explained in \cite{morton11}. In particular, we used the online tool TRILEGAL\footnote{http://stev.oapd.inaf.it/cgi-bin/trilegal} to compute the number of expected stars with a limiting magnitude within 5 degrees-squared centred at the galactic latitude of Kepler-91. We can then integrate $N$ over a certain angular separation (each of which has a particular limiting magnitude in our AstraLux image, see Fig.~\ref{sens}) to compute the total probability of a non-detected background source. We { found that} the background source probability for Kepler-91 { is} below 2.7 \% for our 10\% selection rate AstraLux image. Note that, before our high-resolution image, this probability was (assuming observations and resolutions by the Sloan Digital Survey, SDSS) larger than 7\%. 

On the other hand, we can calculate the probability that a given background source is actually an eclipsing binary  { able to mimic} the signal of a  planetary transit. This probability can be calculated with the correlation explained in equation [14] of \cite{morton11} and { provides} a value of 0.03\% for our target. { In conclusion, } multiplying the probability of a non-detected background source (2.7\%) by the probability of such source being an appropriate binary (0.03\%), we have a $0.0008$\%  { chance} for the existence of a non-detected appropriate eclipsing binary. 

{Although our analysis has reduced the background source probability down to 2.7\%, we have to deal with the possibility of a non-detected blended star. We can measure how a non-detected source with a magnitude difference greater that the $3\sigma$ detection limit in the AstraLux image would affect the planet properties (in particular, the planetary radius). By removing the light contribution of a hypothetical non-detected stellar companion of magnitude difference $\Delta m$, the depth of the transit (and thus the planetary radius) would be increased by the factor given in equation [6] of \cite{lillo-box12}.  For instance, a $\Delta m=6$ mag source at 0.5 arcsec would not have been detected by our high-resolution image (see Fig.~\ref{sens}). As a consequence, the actual transit depth would increase by a factor of 1.0039 (0.39\%) as computed by the aforementioned expression. The last column in Table~\ref{tab:astralux} shows the result of this calculation for each angular separation at a $3\sigma$ detection limit for the $10\%$ selection rate image.}

{ The only relevant configuration that could mimic a planetary transit and cannot be rejected by our high-resolution images is a diluted binary in a triple system. However, \cite{morton11} provided an estimation of the probability for a given transit depth, period and primary mass that such eclipse is produced by a hierarchical triple system. The authors conclude, for a given system with a one solar-mass primary star and a 10 days orbital period (similar to our system) that the probability of such appropriate hierarchical triple system is of the order of 0.001 \% for diluted eclipse depths in the range $10^2-2\times 10^4$ ppm.}

Thus, with these considerations, we assume along this paper that Kepler-91 is isolated and its light curve is not affected by a close companion or any other object along its line of sight.

\subsection{Spectral Energy Distribution analysis \label{sec:SED}}

A zero-order estimate for the stellar parameters of Kepler-91 was obtained using the Virtual Observatory SED Analyzer \citep[VOSA\footnote{http://svo2.cab.inta-csic.es/theory/vosa/}, ][]{bayo08}. Its latest version (Bayo et al. 2013, submitted) uses bayesian inference to compute the expected values for the effective temperature, surface gravity, metallicity and interstellar extinction. We have used { every photometric data-point} available in the literature (to our knowledge) to { build and fit the SED from} Kepler-91. In particular, we used the KIC photometry in the \textit{g, r, i, z} filters \citep{brown11}, the 2MASS JHKs photometry \citep{cutri03}, WISE (Wide-field Infrared Survey) bands W1 to W4 \citep{wright10}, the {\it Kepler} band \citep{borucki10}, and UBV photometry from \cite{everett12}. Table~\ref{tab:photometry} summarizes this information.

The bayesian analysis from VOSA reveals that Kepler-91 has an effective temperature of $T_{\rm eff}=4790\pm110$ K with metallicity being slightly over-solar $\rm{[Fe/H]}=0.4\pm0.2  $ (see summary in Table~\ref{table:hostpars}). We have set the extinction range to $A_V=[0.0,1.0]$ mag. The output expectance and variance from the bayesian probabilities is $A_V=0.43\pm0.15$ magnitude. The surface gravity, however, is not very well constrained but the probability distribution function seems to indicate that $\log g < 3.5  $. These values are in good agreement to the ones obtained by the KIC study \citep{brown11} and \cite{huber13}.

\subsection{Analysis of the high-resolution spectrum\label{sec:spec}}

We used the high-resolution and high-SNR spectrum obtained with CAFE to better constrain stellar parameters and to validate previous values from the SED analysis. In particular we have centred our study in the metallicity and effective temperature values which will be crucial { to better constrain the parameter space } in our own asteroseismic modelling. A previous inspection of the spectrum shows the lack of lithium at 6707.8 \AA, indicating the evolved stage of the host star.

\subsubsection{Metallicity \label{sec:metallicity}}

{ Instead of performing a general fit} to the high-SNR spectrum (which would imply a large number of free parameters), we have performed a { focused} analysis of the metallicity of the star. { The purely} photometric analysis (see section \S~\ref{sec:SED}) provided a value of $\rm{[Fe/H]}=0.4\pm0.2$, an over-solar abundance already determined by previous works

\begin{figure*}[ht]
   \centering
   \includegraphics[width=0.47\textwidth]{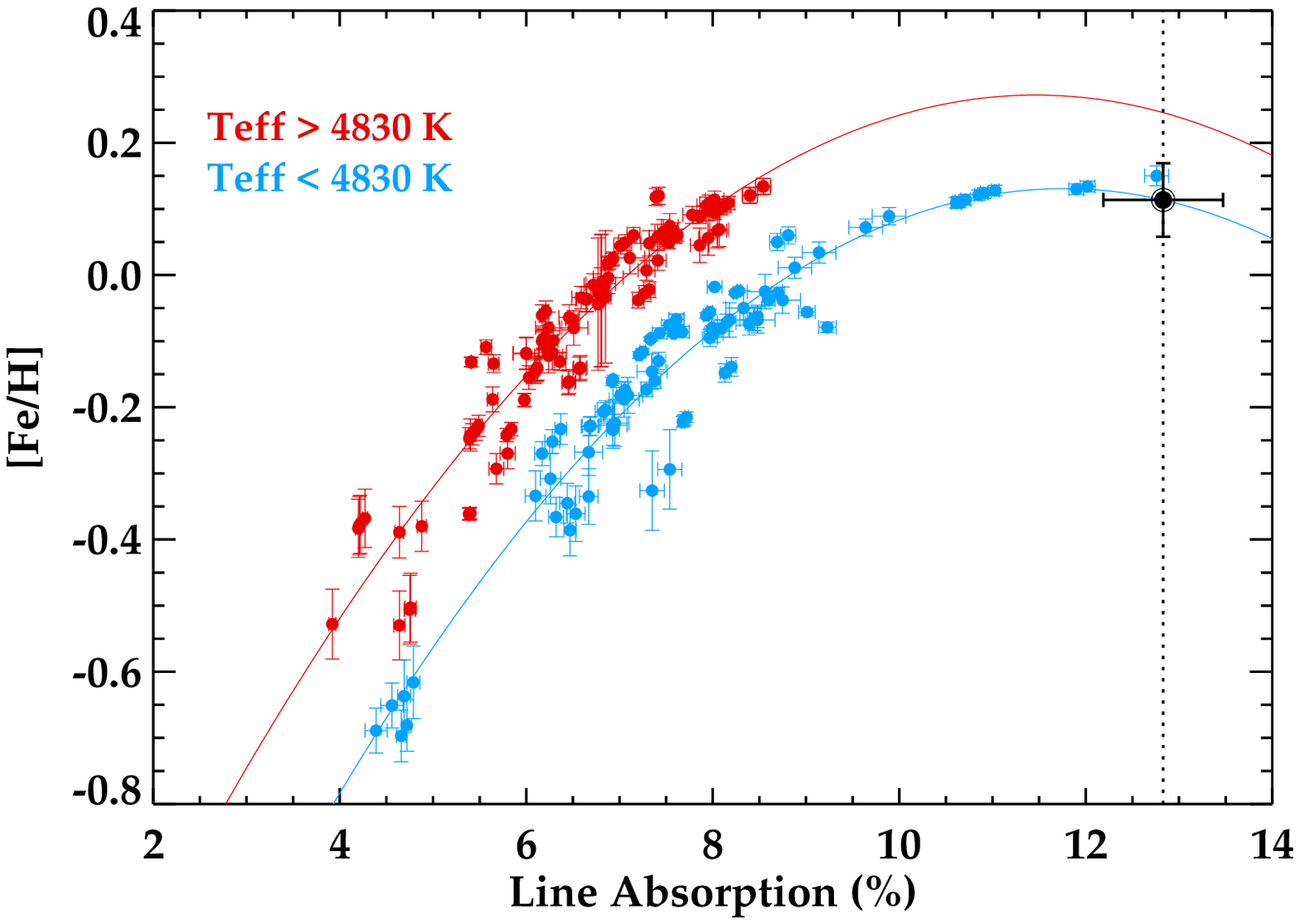} \includegraphics[width=0.47\textwidth]{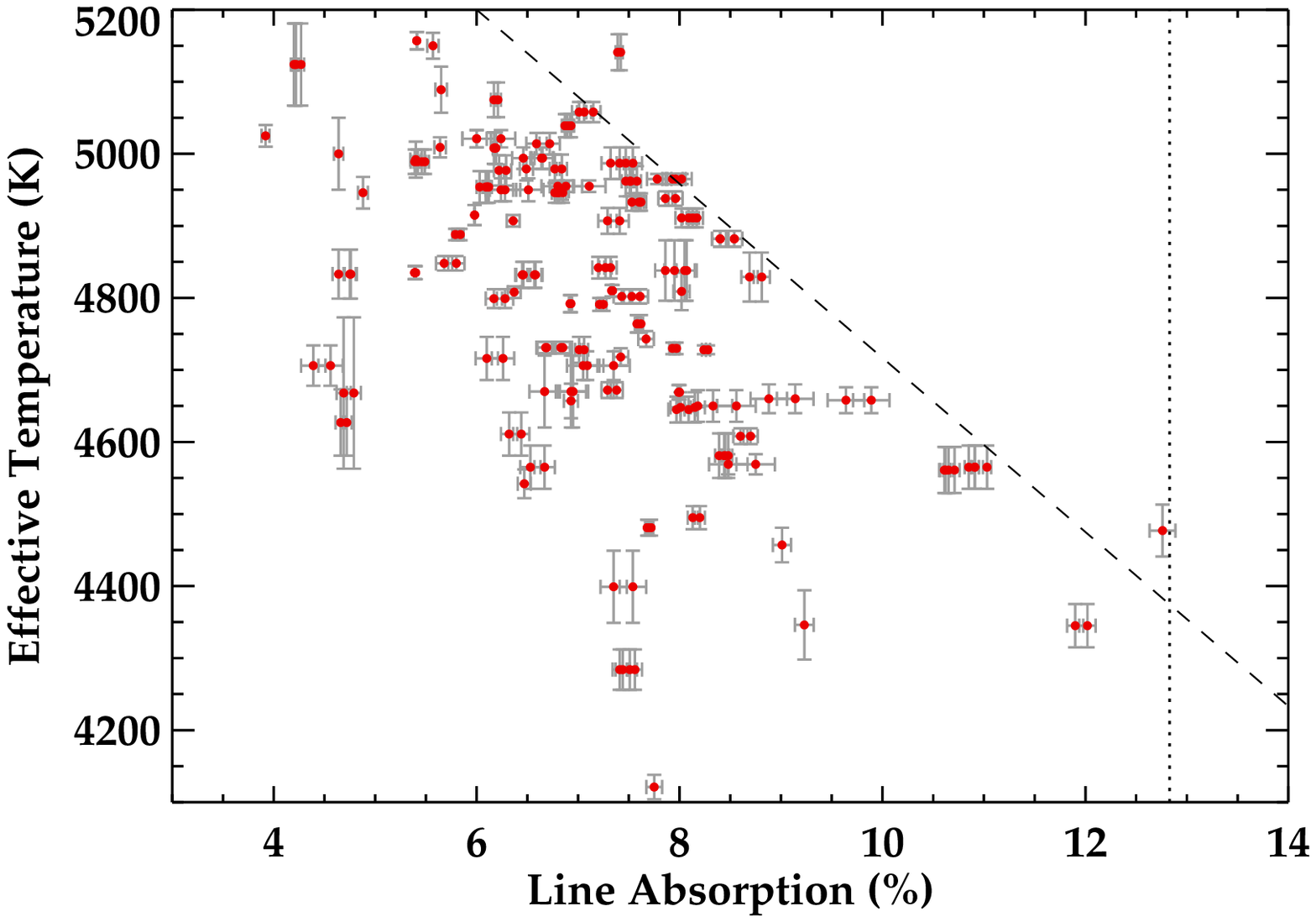}
      \caption{\textbf{Left:}  Determination of the stellar metallicity of Kepler-91 (see section \S~\ref{sec:metallicity}). We show data from giants compiled by \cite{gray02} to parametrize the line absorption parameter (LA) with the iron abundance [Fe/H]. Two samples are shown: giants with $T_{\rm eff}>4830$ K (red) and giants with $T_{\rm eff}<4830$ K (blue). Solid lines represent the fitted second order polynomials to the test data. The dotted vertical line shows the obtained LA for Kepler-91 and the black circle its determined position in the diagram. \textbf{Right:} Effective temperature versus LA showing that Kepler-91 should be considered in the cool group of giants regarding the left panel segregation. The red circles show the position of the test giants and the dashed line represents an estimated empirical limit to the temperature for each LA.}
         \label{fig:metallicity}
   \end{figure*}

We have followed the giant stars specific prescriptions described by \cite{gray02} to obtain an independent value. This scheme uses a small part of the spectrum (from 6219.0 \AA\ to 6261.5 \AA) that was verified to be mainly dependent on the stellar metallicity. The method uses the percentage of stellar continuum absorbed by the atmospheric elements of the star. This percentage is what the authors call the line absorption (LA). After masking specific lines that strongly depend on the effective temperature, they { were} able to { fit} a second order { polynomial that} provides the  value of [Fe/H] for the star as a function of the LA. Since the authors do not provide the coefficients of this polynomial, we used the results in their table 4 to { perform our own fit}. It is important to note the clear (although the { physical} reason is unknown, as the authors claim in their work) difference between stars with $T_{\rm eff}$ above and below 4830 K. We divided the calibration sample into two groups according to this { separation (hot for $T_{\rm eff}>4830$ K and cold for $T_{\rm eff}<4830$)} and fit two different polynomials of the form $\rm{[Fe/H]}=a_0+a_1x+a_2x^2$ with $x$ being the masked line absorption in \%. Coefficients for the fit of both groups are reported in Table~\ref{tab:coefficients}.  In the left panel of Fig.~\ref{fig:metallicity} we have plotted these polynomials together with the tested giant stars in \cite{gray02}, with stars hotter than 4830 K in red and cooler in blue.

We measured the masked LA { for} our spectrum { finding} that $12.8\pm 0.7$ \% of the light coming from the star is absorbed by chemical elements. Due to the important segregation in temperature, we investigated the dependence of the LA with this parameter. The right panel of Fig.~\ref{fig:metallicity} { shows} the values for the test giants. There is a clear desert of such objects in the upper-right region of the figure (high temperatures and high LA values), { highlighted} with { a} dashed line. It is clear from this figure that for the measured LA of Kepler-91 (vertical dotted line), the effective temperature is not expected to lie above 4830 K. Hence, we have used the coefficients corresponding to the cool polynomial to compute { the} metallicity. The uncertainty has been calculated by a quadratic sum of the error of the LA parameter and the standard deviation of the residuals of the test-giants { with} respect to the fitted polynomial\footnote{ This implies that the uncertainty in the metallicity is at $1\sigma$ level. Here we prefer to keep the $1\sigma$ uncertainty to constrain as much as possible the parameter space for subsequent asteroseismic analysis with high computational time consuming.}. { Our final estimation} is $\rm{[Fe/H]}=0.11\pm0.07  $  { which agrees} within the errors, with the one obtained by \cite{huber13}, $\rm{[Fe/H]}=0.29\pm0.16  $.

\subsubsection{Effective temperature \label{sec:teff}}

We have used four line pairs in the spectrum to estimate the stellar effective temperature. { The line depth ratios of these pairs }are used in \cite{gray01} to obtain { temperatures given } the fact that one of the lines is temperature { insensitive} while the second one strongly depends { on} it. We used the pairs Ni{\sc i}/V{\sc i} at 6223.99/6224.51 \AA , Fe{\sc i}/V{\sc i} at 6232.65/6233.20 \AA , V{\sc i}/Fe{\sc i} at 6251.83/6252.57 \AA , and Fe{\sc i}/V{\sc i} at 6255.95/6256.89 \AA. { On a first step, to estimate the rotational velocity, we have synthesized a grid of models using the ATLAS09\footnote{http://kurucz.harvard.edu/grids.html} software for metallicities [Fe/H]=0.0-0.2, effective temperatures in the range 4400-4800 K (50 K step), surface gravities from 2.5 to 3.5, and rotational velocities from 1.0 to 12.0 km/s in steps of 0.1 km/s (turbulence velocity fixed to 2.0 km/s). A global fit to the spectrum provides a posterior distribution for the $v\sin{i}$ parameter with a expectance and variance values of $v\sin{i}=6.8\pm0.2$~km/s. By setting the rotational velocity in the calculated range, considering three values for the surface gravity  ($\log g=$ 2.5, 3.0, and 3.5), and building a finer grid of temperatures with 25 K step,  we proceeded to a least-square analysis of the four line pairs. A bayesian analysis provides the next expectance and variance values for the different gravity values: $4600\pm46$ K for $\log g=2.5  $, $4550\pm47$ K for $\log g=3.0  $, and $4500\pm50$ K for $\log g=3.5$. As a compromise between these values we will adopt an effective temperature of $T_{\rm eff}=4550\pm 75$ K, whose central value nearly corresponds to the surface gravity determined by the asteroseismology \citep[][and our own calculations in next section]{huber13}. }

\subsection{Asteroseismology \label{sec:asteroseismology}  }

\subsubsection{Scaling relations \label{sec:mauro}}
Cool stars with a convective envelope may show solar-like oscillations, that is, pressure oscillation modes stochastically excited by turbulent motions. Their power spectra present a regular pattern modulated by a gaussian shape and are characterized by two global parameters: the frequency at maximum power (thereafter $\nu_{\rm max}$) and the frequency separation ($\Delta\nu$) between consecutive radial order ($n$) modes with the same angular degree ($\ell$). These quantities are linked, via scaling relations ($\Delta\nu \propto \rho^{1/2}$, $\nu_{\rm max}\propto g/T_{\rm eff}^{1/2}$), to global stellar parameters such as total mass, radius and effective temperature \citep{ulrich86,brown91,kjeldsen95,belkacem11}. These relations read:

\begin{equation}
 M_{\star}  = \bigg(\frac{\nu_{\rm max}}{\nu_{\rm max,\odot}}\bigg)^3\bigg(\frac{\Delta\nu_\odot}
{\Delta\nu}\bigg)^4\bigg({\frac{T_{\rm eff}}{T_{{\rm eff},\odot}}}\bigg)^{3/2}
\label{eq:ms}
\end{equation}

\begin{equation}
 R_{\star}= \frac{\nu_{\rm max}}{\nu_{\rm max,\odot}}\bigg(\frac{\Delta\nu_\odot}{\Delta\nu}\bigg)^2\sqrt{\frac{T_{\rm eff}}{T_{{
\rm eff},\odot}}}
\label{eq:rs}
\end{equation}

Note that these equations allow us to derive mass and radius (once we have an estimate of the effective temperature) independently of the chemical composition and of stellar modelling.  They  are, however, approximate relations and  must be calibrated. The validity of  $\Delta\nu \propto \rho^{1/2}$ can be tested with predictions from models,  as done by \cite{white11}, \cite{Miglio2013a}, and \cite{mosser13}. The second relation cannot be tested with models, and only a theoretical justification has been proposed by \cite{belkacem11}. Nevertheless,  comparisons between global parameters derived from seismology and those obtained from interferometry and spectroscopy of solar-like pulsators indicate that $\nu_{\max}$ is a very good proxy of the surface gravity and stellar radius \citep{Miglio2012,morel12,white13,huber12}. These studies suggest that, in the analysed domain, equations ~\ref{eq:ms} and ~\ref{eq:rs} can provide stellar radius and mass with an uncertainty of 4\% and 10\% respectively \citep[][and references therein]{huber13}, and that is a significant improvement with respect to the classical spectroscopic/photometric approach.

These scaling relations are being  extensively used,  in the framework of stellar population studies \citep{miglio09,Miglio2013b,mosser10,mosser11,hekker09,hekker11} and of exoplanet parameter determination \citep[see review by][]{moya11},  to characterize dwarfs and red giants solar-like pulsators detected by CoRoT and {\it Kepler}. In particular, scaling relations have been recently applied to derive the stellar parameters of 66 {\it Kepler} planet-host candidates  presenting solar-like oscillations \citep{huber13}.  Although the information from global parameters $\Delta\nu$ and $\nu_{\rm max}$ is extremely valuable for the study of planetary systems, better and additional constraints (for instance stellar age) can be  expected from individual frequencies.

For the particular case of Kepler-91, \cite{huber13} derived  from the {\it Kepler} light curve the global parameters of the power spectrum:   $\Delta\nu=9.39\pm0.22~\mu$Hz and  $\nu_{max}=108.9\pm3.0~\mu$Hz. We have used the A2Z pipeline \citep{mathur10} to re-determine these values, obtaining $\Delta\nu$ = 9.48 $\pm$ 0.88 $\mu$Hz and $\nu_{max}$ = 109.4 $\pm$ 6.1 $\mu$Hz, in good agreement at 1$\sigma$ level with the previous study, { and leading to a mean density around 2\% larger than that reported in \cite{huber13}. The updated scaling relation suggested by \cite{mosser13}  implies an additional increase of the density of 2\% with respect to that obtained with Eqs. \ref{eq:ms} and \ref{eq:rs}.  According to \cite{mosser13}, these equations should be corrected by a factor of $(1-4\zeta)$ and $(1-2\zeta)$ respectively (with $\zeta=0.038$ for red giants), and the reference values for the Sun should be changed to  $\nu_{\odot} = 3104~\mu$Hz and $\Delta\nu_{\odot} =138.8~\mu$Hz. By using these updated scaling relations and effective temperature ($T_{\rm teff}=4550\pm75$~K), we derive the following  stellar mass, radius, and mean density:  $M_{\star}=1.19^{+0.27}_{-0.22}\,M_\odot$,  $R_{\star}=6.20^{+0.57}_{-0.51}\,R_\odot$, and $\rho_{\star}=7.04\pm0.44$~kg/m$^{3}$ (errors have been calculated by performing Monte Carlo Markov Chain simulations). The corresponding stellar luminosity and surface gravity are: $\log\,g=2.93\pm0.17$, and $L=14.8^{+3.9}_{-3.3}\,L_{\odot}$.}

 Actually, the high signal-to-noise ratio of Kepler-91 power spectrum allows  to detect 38 individual frequencies. In the next sections we try to use them, together with the  spectroscopic results, to  better constrain the properties of this  planetary system.

\subsubsection{Determination of the individual frequencies \label{sec:clara}}

The individual frequencies of Kepler-91 have been obtained by fitting the power spectrum of the signal to a model. For solar-like oscillations the power spectrum shows a $\chi^2$ statistic distribution with two degrees of freedom. Then, a Maximum Likelihood Estimation (MLE) is applied, a method widely used in the determination of p-mode parameters in the Sun and solar-like stars. Following \cite{anderson90}, the likelihood function used for the MLE is:

\begin{equation}
S = \sum [M_i + {O_i \over M_i}]
\end{equation}

\noindent where $O_i$ are the data and $M_i$ is the model, composed of Lorentzian profiles:

\begin{equation}
M_i = \sum {A_i (\Gamma_i /2)^2 \over [({\nu-\nu_i})^2 + (\Gamma_i /2)^2]} +  N(\nu)
\end{equation}

\noindent being $\nu_i$ the oscillation frequency, $\Gamma_i$ the linewidth, $A_i$ the amplitude of each Lorentzian profile, and $N(\nu)$ the noise. $N(\nu)$ is fitted using  two components: constant white noise modelling the photon noise (W), and one Harvey-like profile \citep{harvey85} which reproduces the convective contribution to the background, typically granulation.

\begin{equation}
N(\nu) =  {A \over [1 + ({\nu  /B})^{\alpha}]}  + W
\end{equation}

\noindent where $A$ is related with the amplitude of the granulation, B with its characteristic timescale and $\alpha$ is a positive parameter characterizing the slope of the decay.

The background was fitted prior to the extraction of the modes parameters and then held as a fixed value. For fitting the modes all the parameters are allowed to be free and without any bond among them. The entire spectrum is fitted at once between 65 and 145 $\mu$Hz. The initial values for the p-mode parameters are extracted from the observed spectrum. The formal uncertainties are obtained from the Hessian matrix in the MLE procedure.

\begin{figure*}[ht]
   \centering
   \includegraphics[width=0.55\textwidth]{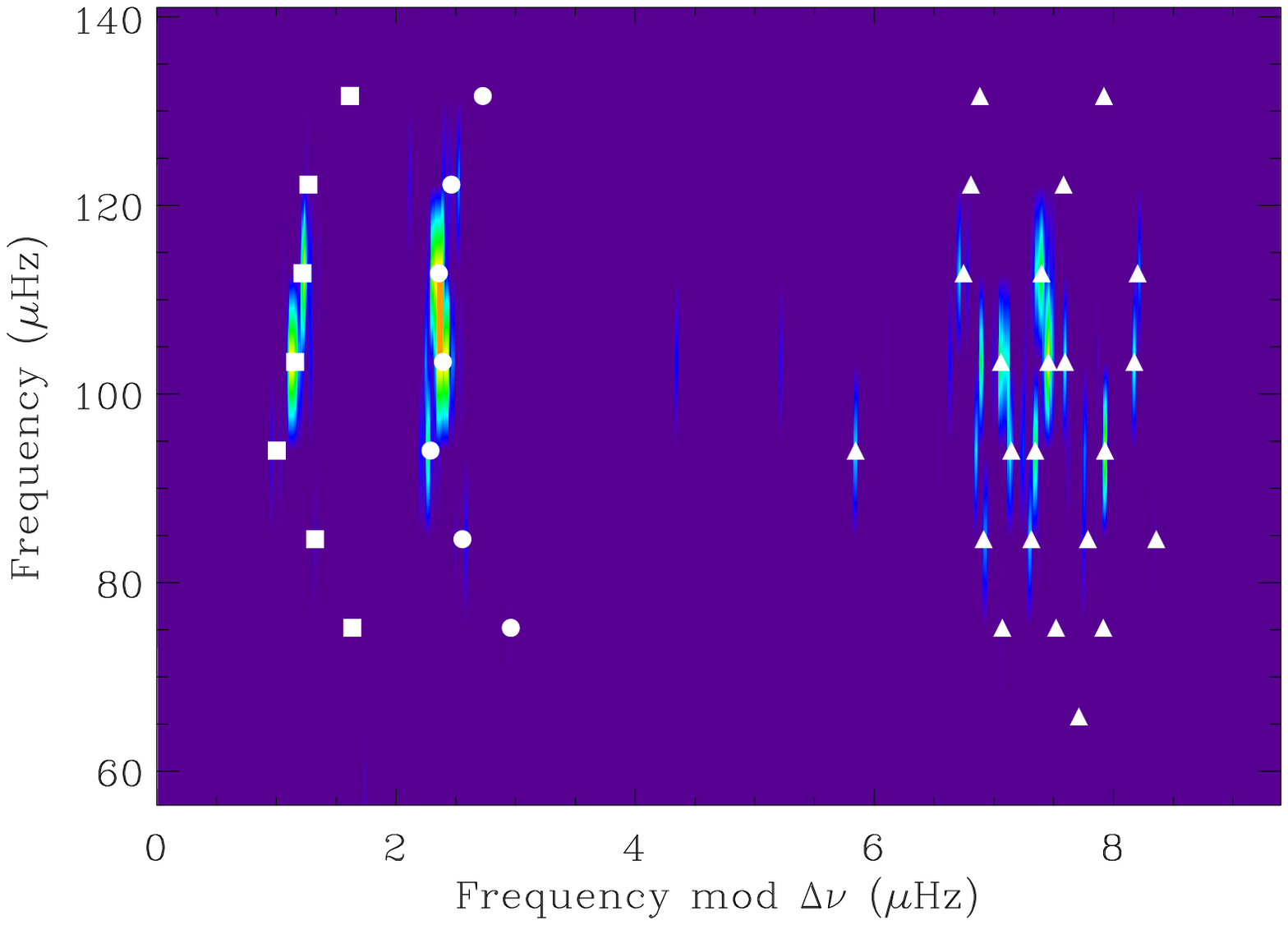}\includegraphics[width=0.45\textwidth]{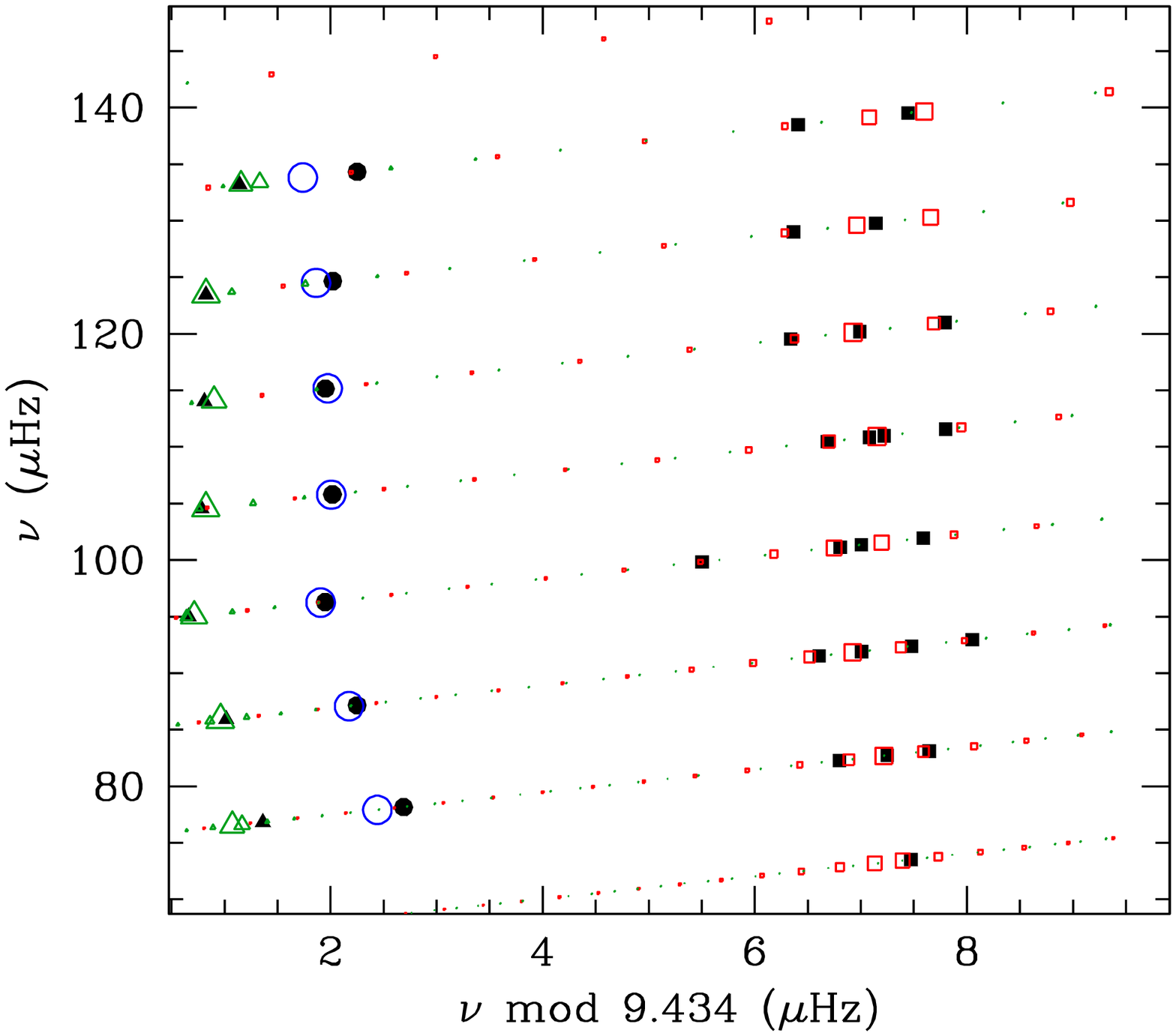}
      \caption{\textbf{Left:} \'Echelle diagram of the power spectrum of the data with the fitted modes overplotted. Circles for $l=0$, triangles for $l=1$ and squares for $l=2$. The power spectrum is fitted using Maximum Likelihood Estimation (see section \S~\ref{sec:clara}).  \textbf{Right:} Comparison between observational (black solid dots and white symbols in the left panel) and theoretical (open symbols) frequencies in the \'echelle diagram for a typical good fitting of radial and non-radial modes. Circles correspond to radial modes, squares to dipole modes and triangles to quadrupole ones. The size of the theoretical symbols is an indication of the expected amplitude based on the value of the inertia mode \citep{houdek99}.  The asymptotic period spacing for this model is 76s. }
         \label{fig:frequencies}
   \end{figure*}

The results are given in Table~\ref{tab:frequencies} and plotted in Fig.~\ref{fig:frequencies} in an \'echelle diagram.

\subsubsection{Individual frequency modelling \label{sec:IndividualModelling}}

The detailed properties of the oscillation modes depend on the stellar structure. In red giant stars, because of the contraction of the inert He core and the expansion of the hydrogen rich envelope, modes  with frequencies in the solar-like domain can propagate in the gravity and acoustic cavities (internal and external regions respectively), presenting hence a mixed gravity-pressure character \citep{dziembowski01,jcd04,dupret09,montalban10}. So, while the solar-like spectra of main sequence pulsators are mainly made of a moderate number of acoustic modes for each angular degree, those of red giants can present, in addition to radial modes, a large number of non-radial g-p mixed modes.

As summarized in Table~\ref{tab:frequencies}, the oscillation spectrum of Kepler-91 presents:  7 radial modes, 7 quadrupole modes, and 23 $\ell=1$ modes. The arrangement of $\ell=2$ and $\ell=0$ modes on well defined vertical ridges in the \'echelle diagram (Fig.~\ref{fig:frequencies}, right panel) suggests that the observed $\ell=2$ modes are well trapped in the acoustic cavity and behave as pure pressure modes. Therefore,  from individual $\ell=0$ and $\ell=2$ frequencies we can derive quantities such as the large and small frequency separations, and try to use them as observational constrains in our analysis \citep{montalban10,bedding10,huber10}. We have computed the mean large frequency separation for radial modes, directly, from frequencies ($\Delta \nu(n,\ell)=\nu_{n,\ell}-\nu_{n-1,\ell}$), and by fitting the asymptotic relation  $\nu_{n\ell}\approx (n+\ell/2+\epsilon)\Delta\nu$ \citep{vandakurov67,gough86,tassoul80}. In the first case we got  $\langle \Delta\nu_0\rangle=9.434\,\mu$Hz with a standard deviation $0.1\,\mu$Hz; and  in the second one $\Delta\nu=9.37\pm0.02\,\mu$Hz. Dipole modes, given their p-g mixed character, do not follow the asymptotic relations for pressure modes,  a certain regularity is expected, however,  in the period spacing between consecutive radial orders \citep{beck11, bedding11, mosser11}, for similarity with the asymptotic behavior of pure gravity modes \citep{tassoul80}.  From the detected dipole modes we got a mean value of the  period spacing of mixed modes ($\Delta P_{obs}$) of the order of 53s. {  This quantity is smaller than the asymptotic period spacing \citep{bedding11,mosser11,dalsgaard12,mosser12,montalban13a} which,  according to \cite{mosser12} estimation for a RGB star with  $\Delta \nu \approx 9.5~\mu$Hz, should be slightly lower than 80s.}

In our fit we also included the spectroscopic constraints, that is $T_{\rm eff}=4550\pm75$~K and [Fe/H]$=0.11\pm0.07$. Taking into account different solar mixtures and the uncertainties in metallicity determination, the constraint used in our fit is then $Z/X=0.019\pm0.005$ ($Z$ and $X$ are the metal and hydrogen mass fractions respectively).

We have used the stellar evolution code ATON \citep{ventura08} to compute a grid of stellar models with masses between 1.0 and 1.6~$M_\odot$ in steps of 0.02~$M_\odot$, helium mass fraction of $Y=0.26-0.32$ in steps of 0.01, metal mass fractions of $Z=0.01$, 0.015, 0.0175, 0.020, and 0.025 and mixing length parameter $\alpha_{\rm MLT}=$1.9, 2.05 and 2.2. The step in radius between consecutive models in the evolutionary tracks is of the order of $5\times 10^{-3}\,R_{\odot}$. For each model with a large frequency separation (from scaling law) within 10\% of the observed value, we compute the adiabatic oscillation frequencies for $\ell=0$, 1, 2 modes using LOSC \citep{scuflaire08, montalban10}. We derived as well  the theoretical values of $\langle \Delta\nu_0\rangle$ and  $\langle\delta\nu_{02}\rangle$.  

The theoretical values of the frequencies and frequency separations differ in general from the observational ones, because of the so-called  near-surface effects. The  model frequencies were  therefore corrected using the method described in \cite{kjeldsen08}.  The power-law correction was applied to radial and non-radial modes. To take into account the different sensitivity of non-radial modes to surface layers, the surface correction  of non-radial modes was multiplied by a factor $Q^{−b}_{n,\ell}$ , where $Q_{n,\ell}$ corresponds to the ratio of the mode inertia to the inertia of the closest radial mode \citep[][chapter 7]{aerts10}. We have considered several values of the exponent $b$ in the surface-correction law: $b=5,6,7, 8$.

For the individual frequency fitting, we have evaluated the agreement between models and observations by using different merit functions (reduced $\chi^2$, $\sum_N ((\nu_{obs} -\nu_{theor})^2/\sigma_{\nu}^2)/N$, including or not the dipole modes). The merit function for radial and quadrupole modes leads to a mean density of $\rho=7.3\pm 0.1$~kg/m$^{3}$.  { This value does not significantly depend on the assumed $b$ parameter in the surface-effects correction. The mean density derived from frequency fitting is therefore 5.8\% larger than that derived from the classic scaling relations (\ref{eq:ms} and \ref{eq:rs}), and 3.7\% larger than that provided by their updated version \citep{mosser13}.  This discrepancy between both methods is in agreement with other studies \citep[see for instance Fig. 4 in][]{belkacem13}. }

 We have also evaluated the fit of the dipole modes in two different ways: one taking into account only the most trapped modes, those with lowest inertia between two radial modes (the largest symbols in the right panel of Fig.~\ref{fig:frequencies}) and another taking into account all the dipole mixed modes. The results obtained by these two methods are consistent, and provide two different minima in the stellar mass-radius domain: one around  1.25~$M_{\odot}$ and the second around 1.45~$M_{\odot}$. The exclusion of solutions with effective temperature deviating by more than  3$\sigma$, reduces the space of parameters to $M_{\star}=1.31\pm0.10$~$M_{\odot}$, $R_{\star}=6.30\pm0.16$~$R_{\odot}$,  $\log g=2.953\pm 0.007$, and an age of $4.86\pm 2.13$~Gyr. 

The frequencies of radial modes varies as $\rho^{1/2}$. Given the steps used in stellar radius and mass we can expect a typical change of frequencies between different models of the order of 0.8\%, that is, 0.9 $\mu$Hz. That value is much larger than the intrinsic precision of the observational frequencies. The computation of non radial frequencies for so evolved object is very time consuming. Moreover, given the uncertainties linked to the surface-effects and its correction \citep[see e.i.,][]{gruberbauer12}, a denser grid of models is not worth.

We have used the theoretical isochrones from \cite{girardi02} to check the self-consistency of the asteroseismicly and spectroscopically determined parameters. We conclude that the results are fully compatible.



\section{Light curve analysis: planetary system parameters \label{sec:primary}}

 In this section we revisit the {\it Kepler} photometric data and analyse the effects on the stellar light curve induced by the object orbiting Kepler-91. In particular, we revise previous solutions and investigate the possibility of a non-circular orbit. We then produce a new transit fitting with the inclusion of the eccentricity ($e$) and the argument of the periastron ($\omega$) as new free parameters. Finally, we will also study other signatures present in the light-curve due to the presence of a close companion. 

For this analysis, we have phase-folded the whole light-curve according to the recently published transit ephemeris and orbital period obtained by \cite{tenenbaum12}. Then, we binned the light-curve to 9 minutes (around 52 original points) with a $3\sigma$ clipping rejection algorithm. No { relevant} improvement is found when { performing a zero-order cleaning of the solar-like oscillations of the star filtering the high frequencies (explained in section  \S~\ref{sec:asteroseismology}) in the Fourier space. Since we will work with more than 200 folded and binned transits, we estimate that these oscillations will be partially masked out and will not play a relevant role.}

\subsection{Revisiting transit parameters \label{sec:newfit}}

The transit of this system has already been previously fitted in the TCE (Threshold Crossing Events) analysis by \cite{tenenbaum12}. The orbital and physical parameters calculated in that paper are summarized in the second column of Table~\ref{tab:REBresults}. However, as it was shown in section \S~\ref{sec:HostProperties}, the physical parameters of the host star are now much better determined. In particular, effective temperature, surface gravity and metallicity are quite different from that assumed by TCE. The dependency of these parameters on the transit shape comes from the limb darkening coefficients. By trilinearly interpolating the \cite{claret11} tabulated values of the four quadratic limb darkening coefficients, we find that the relative differences between adopting the TCE stellar parameters and our determined parameters are of the order of 17\%, 40\%, 20\%, and 7\%, respectively. Thus, a new transit fitting is needed for this system.

    \begin{figure}[ht]
   \centering
   \includegraphics[width=0.5\textwidth]{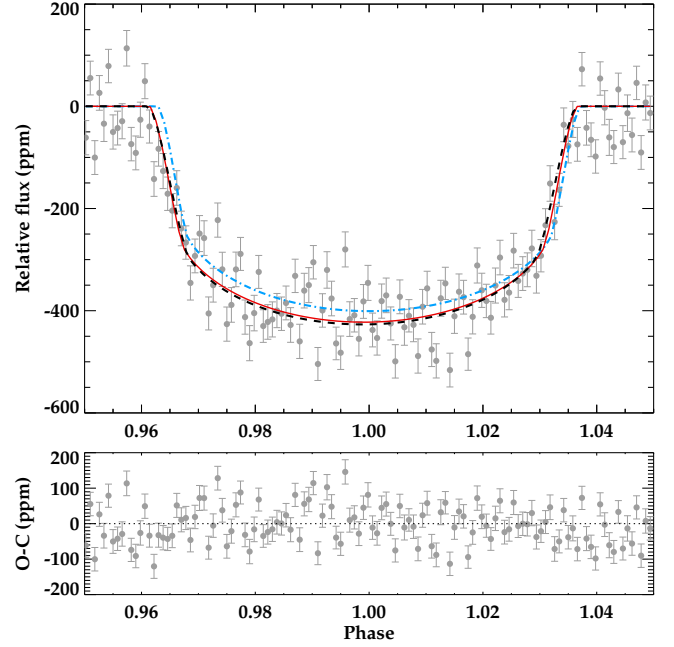}
      \caption{Best-fit solutions for the transit fitting according to different model schemes: assuming parameters from \cite{tenenbaum12} (dot-dashed blue line), assuming zero eccentricity (dashed black line), and leaving the eccentricity as a free parameter (solid red line). Residuals for the fixed $e=0$ model are presented in the lower panel.}
         \label{fig:PrymarySolutions}
   \end{figure}

In TCE, the authors assumed  zero eccentricity for the orbit. The effect of a non-zero eccentricity in the shape of the transit is known to be tight for small values of $e$. It would produce asymmetries in the ingress and egress slopes since the planet-to-star distance at both orbital positions would be different. In order to test this possibility and due to the subtleness of this effect, we have carried out a Kolmogorov-Smirnov test between both sides of the transit. The results show a 70\% of probability for the ingress being equal than the egress (several binnings were tested yielding similar results). Since the magnitude of the asymmetries could be very small and given that there is a non negligible probability of 30\% for the ingress being different than the egress, we find justified to try fitting the transit with a non-zero eccentricity. 

By allowing a non-circular orbit, the transit shape depends on six free parameters: planet-to-star radius ($R_p/R_*$), orbital eccentricity ($e$), argument of the periastron ($\omega$), semi-major axis ($a/R_*$), orbital inclination from the plane of the sky perpendicular to our line of sight ($i$), and  phase offset ($\phi_{\rm offset}$) . This sixth parameter is included to account for possible deviations on the measured time of the mid-transit ($T_0$). Limb darkening coefficients are fixed to the central values of the $T_{\rm eff}$, $\log{g}$ and [Fe/H] since we have checked that, under their confident limits, the quadratic coefficients just vary below 4\%, 6\%, 3\%, and 1\%. Monte Carlo Markov Chain (MCMC) simulations show that these changes are inside the error bars of the final fitted parameters.

We have used a genetic algorithm to model-fit our data (see Appendix~\ref{app:ga}). Due to the large amount of free parameters, we note that the solution is \textit{multi-valuated}. Different sets of solutions equally fit the data, having $\chi^2$ values inside the 99\% of confidence (i.e., presenting differences in the $\chi^2$  value smaller than 16.812 with respect to the $\chi^2_{min}$). Although from statistics we cannot choose a particular set of parameters, we selected the one with the smallest relative errors in all parameters. The parameters of this model are shown in the fourth column of Table~\ref{tab:REBresults}. Errors have been estimated by using 99\% confident contours in $\chi^2$ maps for each pair of parameters. The largest upper and lower errors for all pairs have been used. Interestingly, the selected model has a non-zero eccentricity of $e=0.13\pm0.12$. But, other models inside the 99\% of confidence provide a variety of eccentricities ($e<0.28$), planet-to-star radius ($R_p/R_{\star}$~$\epsilon\ [0.021,0.023]$), semi-major axis ($a/R_{\star}$~$\epsilon\ [2.2,2.8]$, correlated with the inclination parameter), and inclination ($i$~$\epsilon\ [65^{\circ},73^{\circ}]$). It is important to note that, for the calculated stellar radius, all solutions restrict the planet radius to $1.3-1.4\ R_{\rm Jup}$. The $\chi^2$ value for the adopted eccentric model is $\chi^2_{\rm red}=2.86$.

We have also run our genetic algorithm by assuming zero eccentricity, which leaves only four free parameters for the system. In this case, the least relative error solution provides a $\chi^2_{\rm red}=2.86$, and all statistically possible solutions provide parameters within the error bars of this model. 

For comparison purposes, we have also reproduced the model fitted by TCE with their limb darkening coefficients and orbital and physical parameters. This model produces $\chi^2_{\rm red}=3.41$. All three models (TCE, $e=0$, and $e\ne 0$) are plotted in Fig.~\ref{fig:PrymarySolutions}. Both quantitatively (by comparing the $\chi^2$ value) and qualitatively (by inspecting the aforementioned figure), our $e=0$ (fixed) solution improves the quality of the fit from that of TCE. However, to evaluate whether the inclusion of the eccentricity as a free parameter improves or not the fit of the transit, we have used the Bayesian information criterion \citep[BIC, see for example, ][]{schwarz78, smith09}. For a given model solution, the BIC value is calculated as $BIC=N\ln{\chi^2_{min}}+k\ln{N}$, where $N$ is the number of observed points and $k$ is the number of free parameters. A difference greater than 2 in the BIC values of both models indicates positive evidence against the higher BIC value, and a difference greater than 6 indicates a strong evidence. Since $BIC(e=0,{\rm fixed})=624$ and $BIC(e={\rm free})=633$, the eccentric case is not favoured against the zero eccentricity scenario. This means that we do not need the eccentricity to correctly fit the observed transit. However, we have proved that there is a combination of ${e,\omega}$ which also reproduces the transit with similar (inside confident limits) values for the $R_p/R_{\star}$, $a/R_{\star}$ and inclination parameters. Then, we can conclude that the primary transit fitting itself is not enough to determine whether the orbit of the transiting object is eccentric or not.

\subsection{Light-curve modulations: confirmation of a planetary-mass companion}\label{sec:REB}

\subsubsection{Definitions and formulation}

\begin{figure*}[ht]
   \centering
   \includegraphics[width=0.95\textwidth]{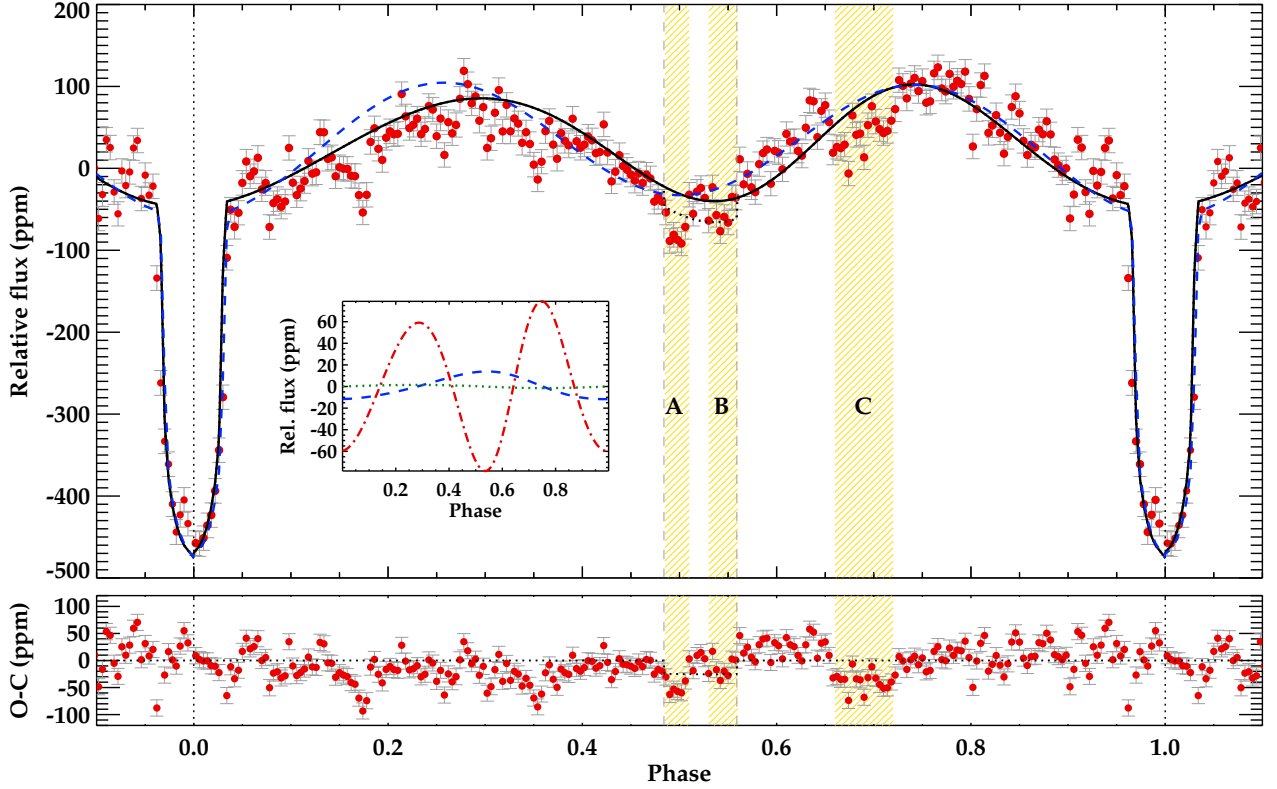}
      \caption{Best-fitted model to the REB modulations for an assumed circular orbit (dashed blue line) and the best model for a non-circular orbit (solid black line). Red circles show the folded light curve assuming the period obtained by the {\it Kepler} team and binned to 29.4 minutes intervals (similar to the real cadence of observations. In the small panel we show the individual contributions of ellipsoidal (red dotted-dashed line), reflection (blue dashed line) and Doppler beaming (green dotted line). The lower panel shows the residuals after substracting the non-circular model to the observed data. The three shadowed regions marked as A, B, and C represent the three detected dims (see section \S~\ref{sec:secondary}). The theoretical location of the secondary eclipse according to the architecture of the system is marked with vertical gray dashed lines in the upper panel. By removing the contribution of the planet reflection in this region, we obtain the dotted black line. }
         \label{fig:REB}
   \end{figure*}

When inspecting the out-of-transit region of the folded light curve (LC) of Kepler-91, a clear double-peaked modulation is apparent (see Fig.~\ref{fig:REB}). This light curve variation is known to be caused by the combination of three main factors in closely packed planetary systems: light coming from the planet (either reflected from the star or emitted by the planet), ellipsoidal variations (or tidal distortions) induced by the planet on the star, and Doppler beaming due to the reflex motion of the star induced by the presence of a massive companion. From now on, we will refer to this LC variations as REB modulations (Reflection, Ellipsoidal, and Beaming). In this section we show the equations and assumptions adopted for the REB fitting in this paper and obtain the solution for the mass of the companion body.

{ Regarding the tidal effect, as stated by \cite{faigler11}, the characterization of the ellipsoidal modulations in non circular orbits is still poorly understood for the case of very low-mass companions orbiting close to the primary star. However, \cite{pfahl08} provided an analytic formula for the excited oscillation modes of the primary component due to a less massive companion. According to it, for systems with null or small eccentricities, the tidal modulation could be modelled by only using the first harmonic ($l=2$) of such oscillations. This harmonic includes a multiplying factor $(a/r)^3$, with $r$ being the planet-star distance given by

\begin{equation}
 r=\frac{a(1-e^2)}{1+e\cos{\psi}} ,
\end{equation}

\noindent where $\psi$ represents the true anomaly. For orbits not aligned with the line of sight, this factor implies different amplitudes at quarter phases. Thus, ellipsoidal variations can serve to constrain the eccentricity ($e$) and the argument of the periastron ($\omega$) of the orbit. The light curve of Kepler-91 shows a small difference in the amplitudes at quarter phases, confirming a small non-zero eccentricity for this system.}

{ According to this considerations, the analytic functions used to fit the observed REB modulations are:}

\begin{equation}\label{eq:ellip}
\frac{\Delta F_{\rm ellip}}{F}=-\alpha_e\frac{M_p}{M_{\star}} ~ \left( \frac{R_{\star}}{a} \right)^3 \left( \frac{1+e\cos{\psi}}{1-e^2} \right)^3 \sin^2{(i)}~ \cos{2\theta} ,
\end{equation}

\begin{equation} \label{eq:doppler}
\frac{\Delta F_{\rm beaming}}{F}=(3-\Gamma)\frac{K}{c} (\sin{\theta} + e\cos{\omega}) , 
\end{equation}

\begin{equation} \label{eq:ref}
\frac{\Delta F_{\rm ref}}{F}=-A_g \left(\frac{R_p}{r}\right)^2 \sin{i} \cos{\theta} , 
\end{equation}

\noindent where $\theta$ represents the angle between the line of sight and the star-planet direction. Its value at each planet position can be obtained from the orbital phase ($\phi$) for a given orbit with eccentricy $e$ and argument of the periastron ($\omega$) by solving the Kepler equations \citep[see equations 3.1.27 to 3.1.34 in ][]{kallrath09}.

{In Eq.~\ref{eq:ellip}, the $\alpha_e$ factor depends on the linear limb darkening coefficients \citep[obtained from trilinear interpolation of the tables provided by][]{claret11} by the expression introduced by \cite{morris85}:}

\begin{equation} \label{eq:alphae}
\alpha_e = 0.15 \frac{ (15+u)(1+g)}{3-u}
 \end{equation}

In the beaming effect, the $\Gamma$ factor is provided in \cite{loeb03}:

\begin{equation} \label{eq:gamma}
\Gamma = \frac{e^x(3-x)-3}{e^x-1}
\end{equation}

\noindent being $x=hc/k_B\lambda T_{\rm eff}$, where we have used $\lambda_{\rm eff}=5750$ \AA\ for the {\it Kepler} band. \\

The parameter $K$ represents the amplitude of the radial velocity, which can be written as:

\begin{equation} \label{eq:k}
 K=28.4 m/s \times \left(\frac{P}{1 yr} \right)^{-1/3} \frac{M_p\sin{i}}{M_{Jup}}  \left(\frac{M_{\star}}{M_{\odot}}\right)^{-2/3} \frac{1}{\sqrt{1-e^2}}
\end{equation}

In the reflection term, the $A_g$ factor represents the geometric albedo of the planet, which was formulated by \cite{kane10} as:

\begin{equation} \label{eq:ag}
 A_g=\frac{e^{r-1}-e^{-(r-1)}}{5(e^{r-1}+e^{-(r-1)})} + \frac{3}{10}, 
\end{equation}

\noindent being $r$ the planet-star distance.

The total light curve modulation can thus be modelled by the sum of all three contributions:

\begin{equation} \label{eq:total}
\frac{\Delta F}{F} = \frac{\Delta F_{\rm ellip}}{F}  +  \frac{\Delta F_{\rm ref}}{F}  +  \frac{\Delta F_{\rm beaming}}{F}
\end{equation}

While the first two contributions were commonly known from the study of stellar binary systems, the Doppler beaming was first detected by \cite{faigler11} \citep[although it was also barely detected by][]{maxted00}. In the era of the high-precision space photometers like {\it Kepler} or CoRoT, these three effects can significantly contribute to the confirmation and characterization  of planet candidates. For example, \cite{quintana13} used observations from {\it Kepler} to independently confirm the hot Jupiter {\it Kepler}-41b via the detection of the REB modulations.

\subsubsection{Fitting the REB modulations \label{sec:REBfitting}}

Among the whole set of parameters involved in Eq.~\ref{eq:ellip} to Eq.~\ref{eq:ag}, some of them can be fixed based on previous sections (see Table~\ref{tab:REBparameters} for a summary of the adopted values). We then have {six} free parameters to fit: eccentricity ($e$), longitude of periastron ($\omega$), planet mass ($M_p$),  semi-major axis to stellar radius ($a/R_{\star}$), inclination ($i$), and phase offset ($\phi_{\rm offset}$).  In the lower part of Table~\ref{tab:REBparameters}, we have constrained the physical limits to the free parameters to restrict the fitting process.

    \begin{figure*}[ht]
   \centering
   \includegraphics[width=1.0\textwidth]{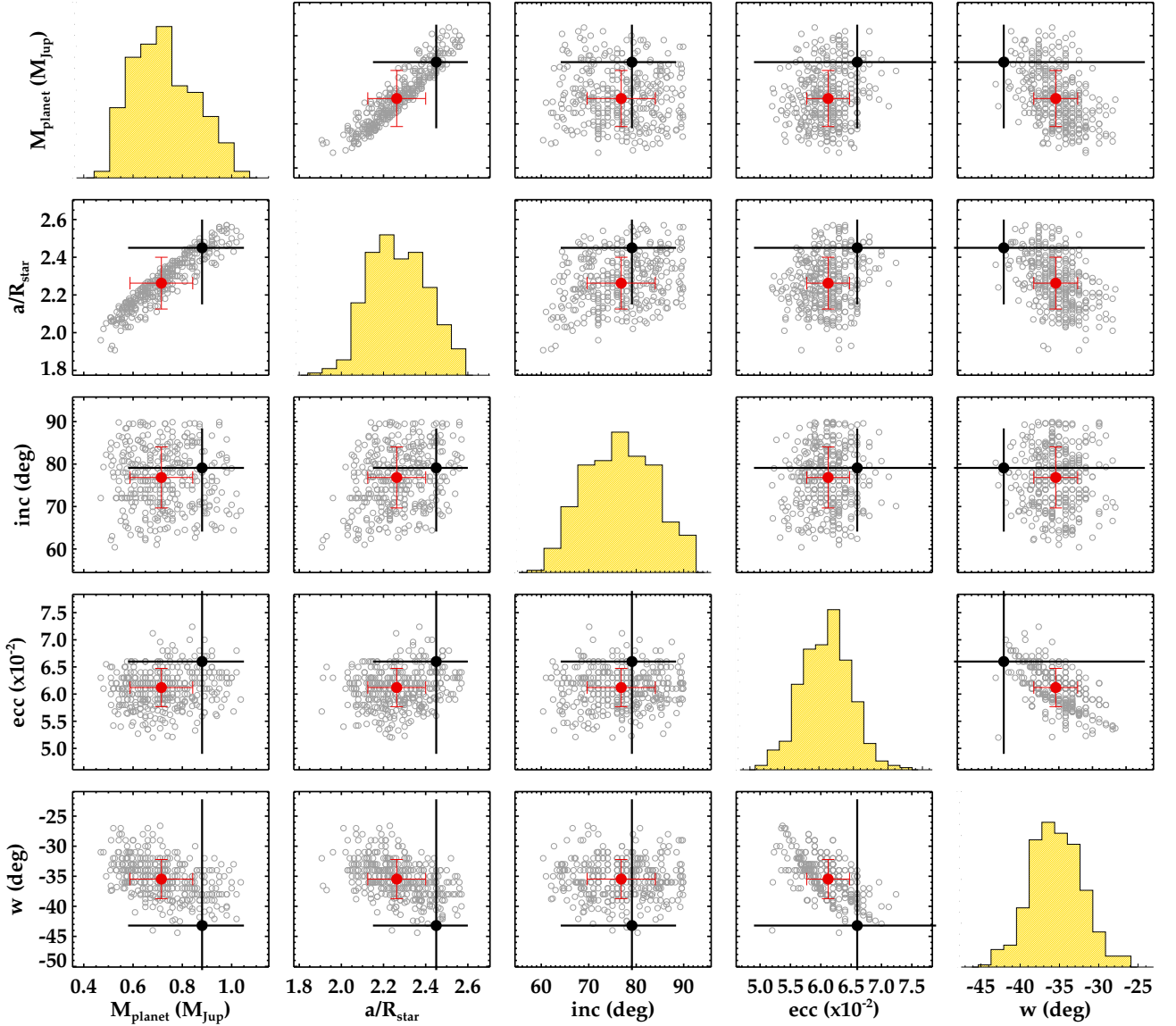}
      \caption{Possible solutions for the REB modulation fitting (see section \S~\ref{sec:REB}). Each grey open circle (and its correspondent error bars) represents a possible solution for the REB modulations whose $\chi^2$ value is statistically valid as compared to the minimum $\chi^2$ obtained with the genetic algorithm. We have marked in black the adopted model, being the one that presents the least relative errors. The red symbol corresponds to the most repeated value (and standard deviation deviation) for each parameter in the whole set of statistically acceptable solutions.}
         \label{fig:PossibleSolutions}
   \end{figure*}

By inspecting the shape of the observed modulation in Fig.~\ref{fig:REB}, key features can be distinguished. The different amplitudes at $\phi=0.75$ could be produced by two causes: either the planet is very massive to produce a large Doppler beaming, or the orbit is actually eccentric and so ellipsoidal variations may not have the same amplitude at quarter phases (as far as $\omega \ne 90^{\circ}$ and $\omega \ne 270^{\circ}$). Since the former effect has its maximum at $\phi=0.25$ and our observations show the maximum just in the opposite peak, we conclude that the difference must be caused by a non-circular orbit.

Since the amplitude of the REB modulations is quite small (100-150 ppm), we have used a larger bin size in this analysis (60 minutes with a $3\sigma$ rejection algorithm). The region where a possible occultation is located has been masked out ($\phi=[0.66,0.72]$). We used our genetic algorithm to model fit the free parameters obtaining 4000 convergence solutions (see detailed explanations about systematic error estimations and the method in Appendix~\ref{app:ga}). {To account for the errors introduced by non-fitted parameters (such as $T_{\rm eff}$, $\log{g}$, $R_p/R_{\star}$, etc.), we have run MCMC simulations allowing these parameters to vary inside their confidence limits. Posterior distributions provide the $1\sigma$ errors, which have been quadratically added to the systematic errors obtained by the genetic algorithm to produce the final parameter errors.}

Since {six} free parameters are fitted, we cannot statistically  disentangle between sets of convergence solutions with a difference in the measured $\chi^2$-value of $\Delta \chi^2=\chi^2 - \chi^2_{min} <16.812$. However, we can choose the model which minimizes the relative errors among the sample of solutions. In particular, this model is also the one that minimizes the error in the companion's mass, a key parameter to confirm its planetary nature. This solution is shown in the last column of Table~\ref{tab:REBresults}. Most importantly, most of the aforementioned possible solutions are contained within the confident limits of these parameters. In Fig.~\ref{fig:PossibleSolutions}, we show the location of all possible solutions in two-dimensional diagrams marking the location of the final adopted value (black symbol) ad the median value for all solutions (red symbol). Note that all common parameters with the transit analysis agree within the error bars, thus providing a \textit{self-consistent} solution for the orbital and physical parameters. { For comparisson purposes, we have estimated a mean amplitude for each modulation assuming the star-planet distance equal to the semi-major axis \footnote{Note that since the best solution provides a non-circuar orbit, the amplitude itself varies with the orbital position of the planet.}. Our derived parameters provide the next peak-to-peak amplitudes: $A_{ellip}=121^{+32}_{-33}$ ppm, $A_{ref}=25^{+15}_{-15}$ ppm, and $A_{beam}=3^{+1}_{-2}$ ppm. As expected, the most relevant effect in this system is the ellipsoidal modulation, given the small separation between the planet and the star.}

{We have also run the fitting algorithm by assuming zero-eccentricity. The best-fit model is shown in Fig. ~\ref{fig:REB} with a dashed blue line. In this case, the improvement in the fit by accounting for non-zero eccentricity becomes clear and could be quantitatively measured by comparing the BIC values of both models: $BIC_{e=0}-BIC_{e\ne 0}=59$. This difference is largely greater than 6 which indicates that the REB modulations are clearly better described by an eccentric model.}

The detection of this ellipsoidal modulation confirms the presence of a physically bound planetary-mass companion to Kepler-91 {\it without} the need for a radial velocity study. It is important to note that all statistically possible solutions mentioned before fit the data with companion masses between 0.5 $M_{Jup}$ and 1.1 $M_{Jup}$, confirming then the planetary nature of the object orbiting Kepler-91. Future radial velocity studies will help to better constrain the planet and orbital parameters.

We find unlikely the possibility of a blended eclipsing binary (not detected by our AstraLux image) as the cause of the ellipsoidal variations measured here. This would require that the background binary orbits the center of masses with the same period as the planet orbits the parent star. Also note that our high-resolution image implied a probability smaller than 2.7\% for a non-detected blended binary, and much less for an eclipsing binary with the specific characteristics needed to mimic these modulations.

\subsection{Detection of other small transits/eclipses \label{sec:secondary}}

Since the orbital parameters have been constrained in previous sections, we can use the \cite{wallenquist50} equation to determine the location of the secondary transit:

 \begin{equation}
 \Delta \phi=0.5 + \frac{e\cos{\omega}}{\pi} (1+\frac{1}{sin^2i})
\end{equation}

\noindent being $\Delta \phi=\phi_{sec}-\phi_{pri}$ the phase difference between primary and secondary eclipses. By doing so and using the orbital parameters from the REB analysis, we get that the secondary eclipse should be centered at $\phi_{sec} = 0.53$. The duration of the secondary eclipse is expected to be similar than that of the primary due to the small eccentricity of the orbit (i.e. around 10-11 hours). The theoretical locus of the secondary eclipse is marked with gray vertical lines in Fig.~\ref{fig:REB}.

After removing the signal produced by the REB variations, three clear dims in the light curve can be detected (see Fig. ~\ref{fig:REB}). The first one is located at the mid orbital period ($\phi_A \approx 0.5$, labelled as $A$ in Fig.~\ref{fig:REB}). Its duration of $d_A\approx 4.5-6.0$ hours is shorter than the primary transit which prevents this from being the secondary eclipse. Secondly, another small dim (labelled as $B$) is found $\phi_B= 0.54$. However, its duration of only 5-6 hours combined with its low signal-to-noise, also prevents us from confirming this as the secondary eclipse.

A third dim ($C$), is found at $\phi\approx 0.68$. In this case, the duration of the possible occultation is of the order of the primary transit's duration ($d_C\approx 11-12$ hours). However, the position of this dim prevents this from being the secondary eclipse of the confirmed planet Kepler-91b. Finally, although not that clear (in shape and location) as the previously analysed dims, there are two more occultations at  $\phi\approx 0.17$ and $\phi\approx 0.35$. Similar reasons as stated for the previous dims discard these other possibilities as the secondary occultation of Kepler-91b.

According to this analysis, we can conclude that none of the previously discussed dims agree with the expected location and duration of the occultation of this system. The contribution of the planet reflection at such orbital phases would yield a theoretical depth of $D_{\rm sec}=25\pm 15$ ppm for the secondary eclipse. Interestingly, this coincides with the depth of the observed dim labelled as B in Fig.~\ref{fig:REB}. However, the theoretical location of the secondary eclipse (shown with gray vertical lines in Fig.~\ref{fig:REB}) encompasses both A and B dims. Thus, although observationally we do not detect a clear secondary eclipse accomplishing all theoretical constraints, we can set an upper limit of 40 ppm for the depth of such eclipse. This would agree with the depth of the two minima labelled as A and B. In other words, the combined position and duration of both dims make plausible, with caveats, the identification of the secondary eclipse. However, more work is needed to unveil the origin of such dims and to confirm the detection of the secondary eclipse.

In any case, the explanation for the three individual dims is beyond the scope of this paper and should be addressed by future work on this planetary system. We have already shown that A and B could be part of the secondary eclipse. Some explanations for the dim labelled as C that should be studied more in detail by future works are listed here:
\begin{itemize}
\item The presence of a large trojan planet (located in the same orbit as Kepler-91b) as the large bodies detected in the L4 and L5 lagrangian points of Jupiter (although stability studies are needed to confirm this possibility), 
\item An outer resonant and transiting planet. This possibility implies non-coplanar orbits since for the measured inclination of $i\approx66^{\circ}$, planets in wider orbits would not transit the parent star.
\item A large exomoon blocking the reflected light from the planet's day side. This configuration would require that the moon's period were an integer number of the planet's period around the host star. Again, this possibility would need an exhaustive stability study.
\item Subtle effects due to the {\it Kepler} reduction pipeline, combined with some activity effect on the stellar surface.
\end{itemize}

Accurate radial velocity measurements for this system would help to more accurately determine the planet and orbital parameters which could feed theoretical studies regarding the stability of the possible explanations for these dims.


\section{Discussion: the planet in context \label{sec:discussion}}

\subsection{Solution for star-planet and orbital parameters \label{sec:final}}

By considering the whole analysis, we report  in Table~\ref{tab:finalresults}  the {  parameters calculated for the Kepler-91 system, and discuss here the most controversial ones.}

\begin{figure*}[ht]
   \centering
   \includegraphics[width=1.0\textwidth]{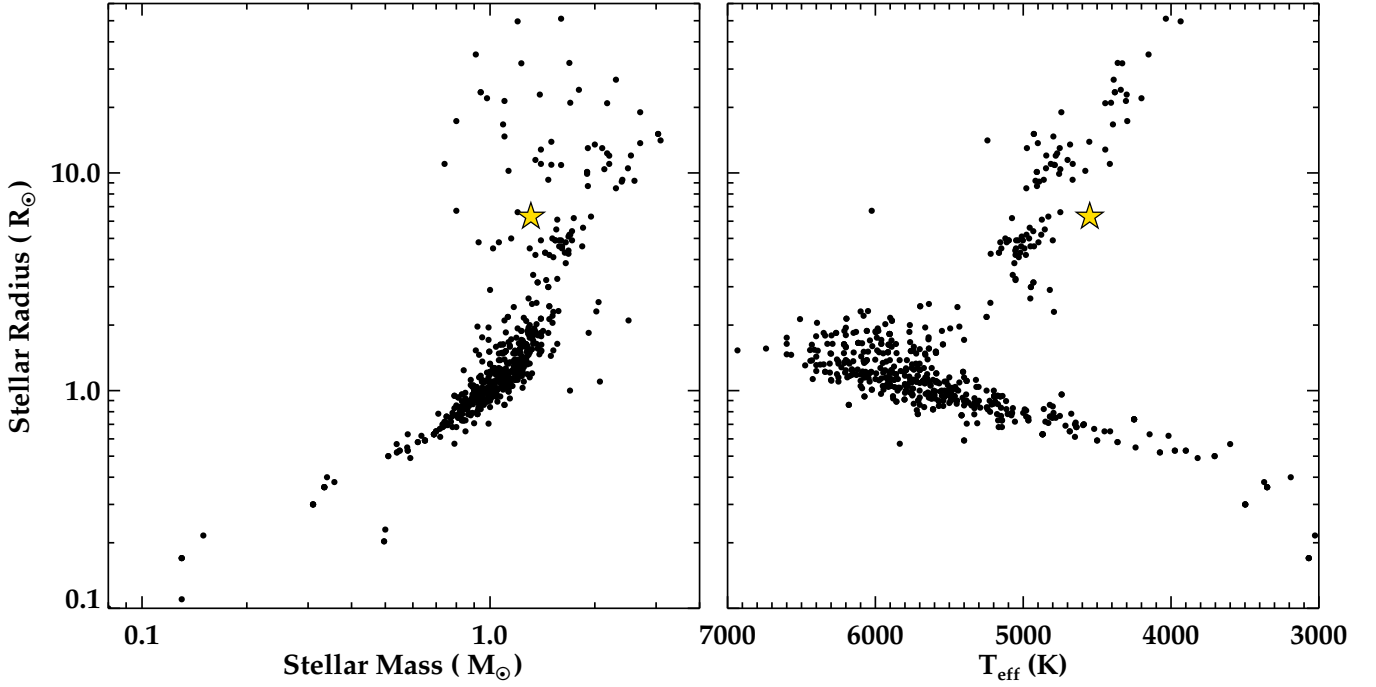}
      \caption{Stellar properties of Kepler-91 compared to other confirmed planetary systems as of July 16th, 2013 (black circles). We have marked with a yellow star-like symbol the location of this system according to the properties obtained in section~\ref{sec:HostProperties}.}
         \label{fig:stellarprop}
   \end{figure*}

Regarding stellar parameters, from all determinations of the effective temperature and metallicity, we have chosen our spectroscopically calculations since they are the most precise ones and lie within the uncertainty limits of other studies. Due to its high precision as compared to other methods, asteroseismology determinations of the rest of the stellar parameters have been assumed. According to both asteroseismology and the analysis of the isochrones and evolutionary tracks, we estimate a stellar age of $2.7-7.0~\rm{Gyr}$. Given the calculated stellar parameters (stellar radius, effective temperature and extinction), we can estimate a distance to Kepler-91 by assuming the bolometric corrections polynomials defined by \cite{flower96} and re-calculated by \cite{torres10}. The calculations provide a value of $d=1030^{+150}_{-130}$~pc. In Fig.~\ref{fig:stellarprop}, {we compare the properties of Kepler-91 to other known stars hosting planetary systems.}

When it comes to the orbital parameters, we have demonstrated that an eccentric orbit is required to better describe the REB modulations. Thus, this parameter (together with the argument of the periastron) is better constrained by this technique rather than by the transit fitting.  Both studies agree in the high inclination of the orbit and provide similar values. We thus use the REB modulation analysis value due to its higher precision.

Finally, given the planetary mass and radius obtained by the REB modulation and transit fitting studies respectively,  we can derive a mean density of $<\rho_p > = 0.33^{+0.10}_{-0.05}~ \rho_{\rm Jup}$,  placing Kepler-91b in the giant gaseous planet locus of the mass-radius diagram of known exoplanets \citep{fortney07}. { As HD209458b, Kepler-91b seems to have an inflated atmosphere probably due to the strong stellar irradiation of its host star. }

\subsection{Comparison to previous works on Kepler-91}

Our results clearly differ with the results from the recently published paper by Esteves et al. (2013) where the planetary nature of Kepler-91b is put into question. The authors base this conclusion on their determination of a geometric albedo greater than 1.0 ($A_g=2.49^{+0.55}_{-0.60}$), corresponding to a self-luminous object and thus a false positive. They use four key parameters to determine this value: their fit to the secondary eclipse, their orbital distance value of $a/R_{\star}=4.51^{+0.12}_{-0.26}$ and the planet-to-stellar radius of $R_p/R_{\star}=0.01775^{+0.00042}_{-0.00065}$ (both coming from the transit fitting), and the assumption of negligible contribution of the thermal emission of the planet. As we have seen, our fits provide quite different values for both parameters, being their semi-major-axis to stellar radius more than twice our value. Since the geometric albedo depends on these parameters in the form:

\begin{equation}
A_g=F_{ecl}\left( \frac{R_p}{a} \right)^{-2} 
\end{equation}

\noindent where $F_{ecl}$ is related to the secondary eclipse depth, we find that the factor multiplying $F_{ecl}$ is more than seven times lower with our determined parameters than for their calculated values. Even assuming that the depth of the secondary eclipse is perfectly fitted (which we have shown to be unclear in section \S~\ref{sec:secondary}), the geometric albedo would then be reduced by a factor of 5.4, placing it below 1 and thus eliminating the self-luminous scenario for Kepler-91b. {  Indeed, the upper limit for the secondary eclipse depth of 40 ppm calculated in section \S~\ref{sec:secondary} implies an upper limit for the geometric albedo of $A_g < 0.5$. Also, if we assume that the theoretical eclipse corresponds to a real detection of the observed eclipse, the calculated depth of $D_{\rm sec}=25\pm 15$ ppm would yield a geometric albedo of $A_g=0.30^{+0.24}_{-0.20}$ for this planet}. The important differences in the physical and orbital parameters are probably due to the almost edge-on orbit calculated by \cite{esteves13}. We have performed an exhaustive fitting of this parameter exploring the whole parameter space and we have found a lower value for the inclination of $i=68.5^{\circ}$ to better fit both the transit and REB modulations (see sections \S~\ref{sec:newfit} and \S~\ref{sec:REBfitting}). We note that such inclination and small orbital distance agrees with the calculations by \cite{tenenbaum12}.

We can also estimate the stellar density (perfectly constrained from asteroseismology to be $\rho_{\star}=7.3\pm0.1~kg/m^3$) directly from photometric observations and fitting as proposed by \cite{seager03} and \cite{sozzetti07}:

\begin{equation}
\rho_{\star}=\frac{3\pi}{GP^2} \bigg(  \frac{a}{R_{\star}}  \bigg)^3  - \rho_p \bigg( \frac{R_p}{R_{\star}} \bigg)^3  
\end{equation}

Although the second term has a negligible influence, we can keep it since we have all the information available. 

If we use the transit fitting results from \cite{esteves13} and their derived planet parameters, we get $\rho_{\star}=44.5 \pm 5.8~\rm{kg/m^3}$, a much larger value than the asteroseismic one. However, our derived parameters from transit and REB modulation fittings provide a much closer estimation of $\rho_{\star}=7.1^{+0.7}_{-1.9}~\rm{kg/m^3}$  { which agrees, within the confident limits with the asteroseismic result. It is important to note that the above expression was derived under the assumption of circular orbit. Thus the result in our case must be taken only as a zero-order estimation and serve only as a comparison to the value computed with the \cite{esteves13} derived parameters.  }

All these considerations together with the fact that they use the photometrically derived stellar parameters by \cite{batalha13}, lead Esteves et al. to compute a companion's mass of 5.92 $M_{\rm Jup}$, if real. Our study uses the more accurate asteroseismic and spectroscopic analysis to determine the stellar parameters,  yielding to a planetary mass of $0.88^{+0.17}_{-0.33}$ $M_{\rm Jup}$ in our final fitting. According to all this information we conclude that Kepler-91b is actually a hot Jupiter planet and that the self-luminous nature proposed by Esteves et al. (2013) is neither clear nor conclusive.

\subsection{Stellar irradiation on the planet}

Due to the extremely close-in orbit of Kepler-91b and the large size of its host, stellar irradiation on the planet's atmosphere should have been playing an important role in the evolution of this planetary system. The equilibrium temperature of the planet is given by \citep{lopez-morales07}:

\begin{equation}
T_{eq}=T_{\star}\left(\frac{R_{\star}}{a} \right)^{1/2}  [f(1-A_B)]^{1/4}
\end{equation}

    \begin{figure}[ht]
   \centering
   \includegraphics[width=0.5\textwidth]{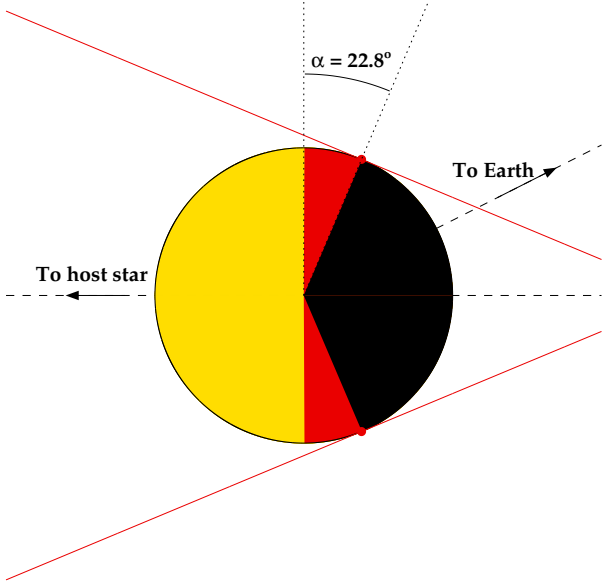}
      \caption{Diagram illustrating the irradiation of the host star onto the planet surface at mid-transit. The red lines represent the boundaries of the stellar irradiation that hits the planet surface.The yellow part represents the dayside if the planet would be farther away. The black part represents the night side and the red one is the extra region illuminated due to the close-in orbit and the large stellar radii.}
         \label{fig:irradiation}
   \end{figure}

{ \noindent where $A_B=\frac{3}{2} A_g$ is the Bond albedo if we assume Lambert's law. According to Eq.~\ref{eq:ag}, the geometric albedo for the calculated parameters is $A_g=0.154$. The $f$ parameter describes the redistribution of the incident stellar flux in the planet's atmosphere and goes from $f=1/4$ when the energy is instantaneously redistributed in the atmosphere to $f=2/3$ when the energy is instantaneously re-radiated to space. The equilibrium temperature in both extreme cases would be: $T_{eq}(f=2/3)=2460^{+120}_{-40}$ K and $T_{eq}(f=1/4)=1920^{+100}_{-30}$ K}. Note that neither \cite{esteves13} nor us have taken into account the effect on the planet equilibrium temperature due to the actual fraction illuminated by the host star, as discussed in \cite{guillot10}. Given the calculated stellar and orbital parameters, we obtain that around 70\% of the planet atmosphere would be illuminated by the host star (in contrast to the approximately 50\% illuminated for planets on wider orbits). The extra angle illuminated of the planet as counted from the perpendicular axis to the orbital plane is, at the periastron (apastron) of the orbit, $\alpha=22.7^{+1.7}_{-1.1}$ ($20.8^{+2.3}_{-1.0}$) degrees (see Fig.~\ref{fig:irradiation}). The angular size of the star as seen from the planet is given by $\tan{(\beta/2)}=R_{\star}/a$ so that, for this system, we obtain $\beta=46.5^{+3.4}_{-0.3}$ degrees at pericenter passage. This value is well above the rest of the known planetary systems ($\beta<10^{\circ}$). The implications of this effect on the determination of the planet and orbital parameters must be investigated by future works.

\subsection{A giant planet at the end of its life}

\begin{figure}[ht]
   \centering
   \includegraphics[width=0.5\textwidth]{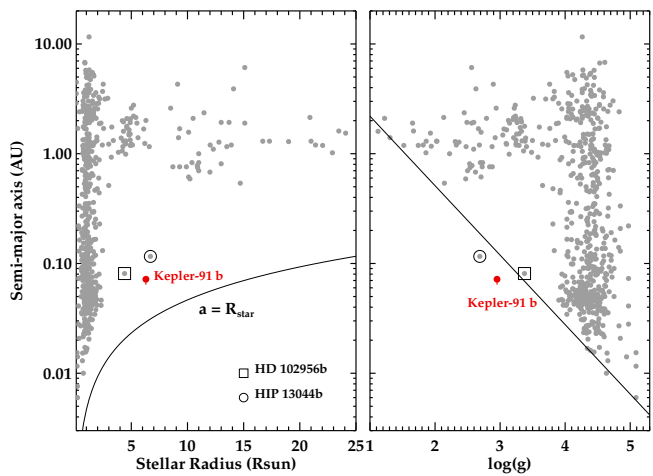}
      \caption{Properties of the Kepler-91 system from a comparative perspective. We plot the semi-major axis of all confirmed planets (grey circles) against the stellar radius (left) and the surface gravity (right). In the left panel, the red point marks the position of Kepler-91 and the solid line is the location of the stellar radius (1:1 line), showing that, among giant stars, this is the closest planet to the stellar surface. In the right panel, we can check that Kepler-91 lies below the solid line marking the empirical limit suggested by Nowak et al. (2013). Together with the intriguing case of HIP 13044, KOI2133 is the only planet that goes through this line.}
         \label{fig:singularity}
   \end{figure}

According to the derived  host star properties in section \S~\ref{sec:HostProperties} and orbital parameters of the planet in sections \S~\ref{sec:newfit} and \S~\ref{sec:REB}, Fig.~\ref{fig:singularity} highlights the singular location of this system. We plot the semi-major axis of all confirmed planets (grey circles) against the stellar radius and the surface gravity. On the left, the red point marks the position of Kepler-91 and the solid line is the location of the stellar radius (1:1 line), showing that, among giant stars, this is the closest planet to its host star. The cases of HD 102956b \citep{johnson10} and HIP 13044b \citep{klement10} are marked as a reference. On the right, we can check that Kepler-91 lies below the solid line marking the empirical limit suggested by \cite{nowak13}. Together with the intriguing case of HIP 13044, Kepler-91 is the only planet that goes beyond this line.

We have used evolutionary tracks from \cite{girardi02} and assumed the effective temperature, metallicity, stellar radius, and mass from our spectroscopic and asteroseismic studies to compute the time that the radius of the star Kepler-91 will reach the current planet's orbital pericenter. If we only take into account this evolutive constraint, we conclude that Kepler-91b will be engulfed by the stellar atmosphere in less than 55 Myr. It is important to note that other non-negligible effects inducing instabilities on the planetary orbit could speed up the planet engulfment. Hence, this result can be considered as an upper limit to the planet's life. The first clear evidences of planet engulfment were published by \cite{adamow12}. The authors showed signs of a post-planet engulfment scenario for BD+48 740, where the presence of a highly eccentric ($e=0.67$) secondary planet and an overabundance of lithium in the stellar spectrum could be caused by the previous engulfment of an inner planet. With a similar stellar mass,  Kepler-91 could be in the immediately previous stage of BD+48 740, the scenario before the planet engulfment.

The planet-to-star distance measured in this paper ($a/R_{\star}=2.45^{+0.15}_{-0.30}$ equivalent to $0.072^{+0.002}_{-0.007}$ AU if we assume the calculated stellar radius) places Kepler-91b as the closest planet to a host giant star. From the sample of confirmed exoplanets\footnote{http://exoplanet.eu/}, none of them is closer than $a/R_{\star}=3.0$. Our derived orbital distance implies that more than 8\% of the sky as seen from the planet is covered by the star (compared to the 0.0005\% covered by the Sun from the Earth). This reinforces the idea that the planet under study is at the end of its life, with its host star rapidly inflating while ascending the RGB. Since we have found no overabundance of lithium in the spectrum of the host star, we conclude that the planet is still not being evaporated (at least the material has not been incorporated into the stellar atmosphere) and that no previous engulfment seems to have happened. 

The non-zero eccentricity for such close-in planet is { intriguing}. { Several effects can produce eccentric orbits of relatively close-in planets \citep[e.g., other non-detected objects, tidal interactions or instabilities due to the mass-loss in the transition from the main-sequence to the giant stage of the host, ][]{debes02}. Thus, a deeper study in this line is needed to unveil the origin of the eccentric orbit of this planet.}

\section{Conclusions \label{sec:conclusions}}

{ In this paper we have proved the planetary nature of the object orbiting the K3 giant star Kepler-91 ($R_{\star}=6.30\pm 0.16~R_{\odot}$, $M_{\star}=1.31\pm0.10~M_{\odot}$). The modeling of the transit signal and light curve modulations allowed us to measure a planetary mass and radius of $M_p=0.88^{+0.17}_{-0.33}~M_{\rm Jup}$ and $R_p=1.384^{+0.011}_{-0.054}~R_{\rm Jup}$. The planet encompasses its host star  in a very close-in eccentric orbit ($a=2.45^{+0.15}_{-0.33}~R_{\star}$, $e =0.066^{+0.013}_{-0.017}$). }

{ We have determined that the expansion of the outer atmospheric layers of the giant star will reach the pericenter of the orbit in less than 55 Myr. The strong stellar irradiation induced by the close star is probably the cause of the inflated atmosphere presented by Kepler-91b and reflected in its low density ($<\rho_p > = 0.33^{+0.10}_{-0.05}~ \rho_{\rm Jup}$). The combination of a large stellar radius and a very close-in orbit, implies that more than 36\% of the day-sky of the planet is covered by the star. This combination also implies that around the 70 \% of the planet atmosphere is illuminated by the host stars. Detecting such additional illumination represents a challenge for the current and future high precision photometers.}


\begin{acknowledgements}
      This research has been funded by Spanish grants AYA 2010-21161-C02-02, AYA2012-38897-C02-01, and PRICIT-S2009/ESP-1496. J. Lillo-Box thanks the CSIC JAE-predoc program for the PhD fellowship. We also thank Calar Alto Observatory, both the open TAC and Spanish GTO panel for allocating our observing runs, and the CAHA staff for providing such useful information about the new \'echelle spectrograph. This publication makes use of VOSA, developed under the Spanish Virtual Observatory project supported from the Spanish MICINN through grant AyA2008-02156. A. Bayo ackowledges funding and support from the Marie Curie Actions of the European Commission FP7-COFUND. B. Montesinos is supported by spanish grant AYA2011-26202. J. Montalb\'an acknowledge financial support from Belspo for contract PRODEX-CoRoT.

\end{acknowledgements}


\begin{appendix} 

\section{Genetic Algorithm \label{app:ga}}

A correct sampling for a model depending on large number of free parameters would yield a huge (unmanageable) grid of models. For instance, in a case where observations could be modelled with six parameters, by sampling each parameter with 15 values,  that would generate $6^{15}$ different sets of parameters. Depending on the time employed by a computer to calculate each model, the amount of time could last for weeks for a good sampling. Instead, a clever way to select the particular set of parameters to be tested can save us a lot of time. This is actually the main goal of the so-called genetic algorithms (GA). 


We have written a genetic algorithm (IDL-based) to explore the whole parameter space without the need of creating a model grid. Basically, a range (minimum and maximum) for each parameter must be supplied, and optionally a prior (initial guessed) value for each parameter. The program performs the following steps: 
\begin{enumerate}
\item If provided, it uses the prior values as parents for the zero-age generation of individuals  (i.e., sets of parameters).  
\item It generates a random population of N individuals within the range limits set by the user and based on the parents in step 1, modifying their values from -20\% to 20\% of the total range, with a certain probability (also provided by the user but typically of the order of 10-30\%). This allows some of the parameters to change while other remain with the same values.
\item It generates the N models with the corresponding parameters. In order to get these models, the user has to provide the function to this end. For instance, in the case of transit fitting, one should provide a function which calculates the \cite{mandel02} model for a given set of parameters or individuals.
\item It obtains the $\chi^2$ value for each individual comparing the models calculated in (3) to the provided data-points.
\item It selects the best 10\% individuals and use them as parents for the new generation to be created in step (1). 
\end{enumerate}

The process is reiteratively repeated until a stable point is reached. This occurs when, after a specified number of generations, the best $\chi^2$ does not change more than 1\% with respect to the best value of the previous generation. At this point, a random population of $10\times N$ new individuals is added to destabilize the minimum achieved to avoid local minima. If the stable point remains, we start again steps (1)-(5) until we reach again a stable point. The process is then repeated 10 times. Hence, the final value has remained stable during more than 100 generations, with 10 inclusions of $10\times N$ random points in the whole parameter space.

Once the final value has been obtained, we proceed to obtain the errors. We use generated $\chi^2$ maps for each pair of parameters centred on the best-fit parameters obtained by the GA. We slightly modify the GA values of each pair while keeping the others with the best-fit values. Then we measure the goodness of the fit ($\chi^2$). Finally, we compute $\chi^2$ contours for each pair for values $\chi^2=\chi^2_0+ \Delta$, where $\Delta$ is a tabulated value depending on the number of free parameters and the confidence limits to be achieved. We used 99\% confident limits to ensure a good determination of the uncertainties in our parameters. Finally, if no correlation between parameters is found (the so-called banana-shaped contours), we select the upper and lower errors for each parameter as the largest ones obtained for all pairs.

 


\end{appendix}

\clearpage


\begin{table*}[h]
\setlength{\extrarowheight}{3pt}
\caption{\label{tab:astralux}  Sensitivity limits (in magnitude difference between the target and a theoretical companion) achieved in the $i_{SDSS}$ band of the Calar Alto 2.2m/AstraLux high-resolution images for 4 selection rates.}
\centering
\begin{tabular}{cc|cc|cc|cc|ccc}
\hline\hline

Ang.Sep.  & Sep.\tablefootmark{a} & \multicolumn{2}{c|}{1\%} &  \multicolumn{2}{c|}{2.5\%} & \multicolumn{2}{c|}{5\%} & \multicolumn{2}{c}{10\%} & $\Delta F_{new}/\Delta F_{old}$ \\
  arcsec & AU & $3\sigma$ &  $5\sigma$  &  $3\sigma$ &  $5\sigma$  &  $3\sigma$ &  $5\sigma$  &  $3\sigma$ &  $5\sigma$ &  upper lim. \\ \hline
       0.20 &$  206^{+  30 }_{-  26 }$  &  0.00   &   0.00  &   0.00   &   0.00 &   0.06  &     --- &   0.03  &   ---    &  1.974 \\
       0.30 &$  309^{+  45 }_{-  39 }$  &  1.06   &   0.11  &   0.94   &   0.00 &   1.12  &    0.13 &   1.10  &   0.08   &  1.362 \\
       0.40 &$  412^{+  60 }_{-  52 }$  &  1.56   &   0.59  &   1.47   &   0.53 &   1.58  &    0.59 &   1.51  &   0.52   &  1.248 \\
       0.50 &$  515^{+  75 }_{-  65 }$  &  2.85   &   1.09  &   2.29   &   1.00 &   2.30  &    1.01 &   2.03  &   0.91   &  1.154 \\
       0.60 &$  618^{+  90 }_{-  78 }$  &  4.14   &   1.75  &   3.75   &   1.47 &   3.66  &    1.55 &   3.29  &   1.40   &  1.048 \\
       0.70 &$  720^{+ 100 }_{-  90 }$  &  4.79   &   2.78  &   4.44   &   2.44 &   4.30  &    2.40 &   4.02  &   2.11   &  1.025 \\
       0.80 &$  820^{+ 120 }_{- 100 }$  &  5.11   &   3.16  &   4.96   &   3.04 &   4.85  &    2.99 &   4.58  &   2.83   &  1.015 \\
       0.90 &$  930^{+ 130 }_{- 120 }$  &  5.48   &   3.55  &   5.27   &   3.38 &   5.23  &    3.38 &   5.05  &   3.28   &  1.010 \\
       1.00 &$ 1030^{+ 150 }_{- 130 }$  &  5.54   &   3.74  &   5.62   &   3.76 &   5.58  &    3.73 &   5.42  &   3.63   &  1.007 \\
       1.10 &$ 1130^{+ 170 }_{- 140 }$  &  5.64   &   3.99  &   5.76   &   4.02 &   5.81  &    4.01 &   5.76  &   3.96   &  1.005 \\
       1.20 &$ 1240^{+ 180 }_{- 160 }$  &  5.64   &   4.18  &   5.86   &   4.25 &   5.96  &    4.28 &   6.00  &   4.27   &  1.004 \\
       1.30 &$ 1340^{+ 200 }_{- 170 }$  &  5.64   &   4.33  &   5.85   &   4.42 &   6.03  &    4.52 &   6.13  &   4.52   &  1.004 \\
       1.40 &$ 1440^{+ 210 }_{- 180 }$  &  5.58   &   4.39  &   5.81   &   4.55 &   6.03  &    4.69 &   6.20  &   4.74   &  1.003 \\
       1.50 &$ 1550^{+ 230 }_{- 200 }$  &  5.54   &   4.46  &   5.76   &   4.64 &   5.94  &    4.78 &   6.14  &   4.87   &  1.004 \\

\hline
\hline

\end{tabular}

\tablefoot{ We have assumed the distance measured in section \S~\ref{sec:SED} to compute the physical separations in the second columns. The last column illustrates the factor by which the transit depth (and thus planet radius) would be modified according to the $3\sigma$ sensitivity limits of a 10\% selection rate.\\
\tablefoottext{a}{Separation in Astronomical Units (AU) assuming a distance of $1030^{+150}_{-130}$ pc for Kepler-91 (see section \S~\ref{sec:final})}
}
\end{table*}


\begin{table*}[h]
\caption{Summary of the results for the host star properties from the different methods explained in section  \S~\ref{sec:HostProperties}. }            
\label{table:hostpars}      
\centering                          
\begin{tabular}{r | c c c c c c c c c }        
\hline\hline 
Method (section)                       &  $M_*(M_{\odot})$     & $R_*(R_{\odot})$ & $\log{g}$      & $\rho~$(kg/m$^3$) &  [Fe/H]\tablefootmark{b}   & $T_{\rm eff}$                & Age (Gyr) & $L_* ~(L_{\odot})$ \\   \hline 
                              &                       &                  &               &                 &                        &                          &           &  \\
KIC10 (\S~\ref{sec:AncillaryPars})           & 1.45                   & 7.488                  & $\mathbf{2.852\pm0.5}$                          &$4.86$           &  $\mathbf{0.509\pm0.5}$              &  $\mathbf{4712\pm200}$                   & N/A                                   &   N/A   \\  [5pt]
TCE (\S~\ref{sec:AncillaryPars})             & 1.49                   & 7.59                   & $\mathbf{2.85}$                                   &$4.80$           &  (-0.2)\tablefootmark{a}      & $\mathbf{4837}$                            & $2.66\pm0.83$ & N/A\\  [5pt]
Huber13 (\S~\ref{sec:AncillaryPars})         & $1.344\pm0.169$                         & $6.528\pm 0.352$       & $2.94\pm0.17$                                   &$6.80$           & $\mathbf{0.29\pm0.16}$               &  $\mathbf{4605\pm97}$                    & N/A                                   & N/A \\ [5pt]
SED (\S~\ref{sec:SED})               & N/A                                     & N/A                    & $ <3.5$                                & N/A                              & $\mathbf{0.4\pm0.2}$                 & $\mathbf{4790\pm110}$                    & N/A                                   & N/A  \\[5pt]            
Spec. (\S~\ref{sec:spec})               &   N/A                                   &    N/A                 & $\mathbf{3.0\pm0.3}$                            & N/A                              & $\mathbf{0.11\pm 0.07}$              &  $\mathbf{4550\pm75}$                    & N/A                                   &  N/A \\ [5pt]
Sc.Rel. (\S~\ref{sec:mauro}) & $1.19^{+0.27}_{-0.22}$                  & $6.20^{+0.57}_{-0.51}$        & $2.93\pm 0.17$                 &   $\mathbf{7.0\pm0.4}$                  &  N/A                        &    $(4550\pm75)$\tablefootmark{a} & N/A & $14.8^{+3.9}_{-3.3} $   \\  [5pt]
Freq. (\S~\ref{sec:IndividualModelling}) & $\mathbf{1.31\pm 0.10}$                  & $\mathbf{6.30\pm 0.16}$        & $\mathbf{2.953\pm 0.007} $                 &   $\mathbf{7.3\pm 0.1}$                  &     $(0.11\pm 0.07)$\tablefootmark{a}      &    $(4550\pm75)$\tablefootmark{a} & $\mathbf{4.86\pm 2.13}$ & $\mathbf{16.8\pm1.7}$    \\  [5pt]


\hline 
\end{tabular}
\tablefoot{\\
Parameters in bold represent primary values (i.e., a directly determined parameter by the used method).\\
Values neither in bold nor in brackets have been calculated based on other previously determined or assumed parameters. \\
The expression N/A reflects parameters that cannot be determined by the correspondent method.\\
\tablefoottext{a}{Assumed (input) parameter, also in parenthesis.}\\
\tablefoottext{b}{Note that $[M/H] \approx log(Z/Z_{\odot}$)}\\
}
\end{table*}

\begin{table*}[h]
\setlength{\extrarowheight}{3pt}
\caption{\label{tab:photometry}  Photometric data used in our spectral energy distribution fitting of section \S~\ref{sec:SED}}
\centering
\begin{tabular}{lcl}
\hline\hline

Band & Magnitude\tablefootmark{a} & Reference \\ \hline

U   &  $15.297 $         & \cite{everett12}\tablefootmark{b}           \\
B   &  $13.986 $         & \cite{everett12}\tablefootmark{b}           \\
V   &  $12.884 $         & \cite{everett12}\tablefootmark{b}           \\
g'  &  $13.407 $         & KIC, \cite{brown11}\tablefootmark{c}        \\
r'  &  $12.406 $         & KIC, \cite{brown11}\tablefootmark{c}        \\
i'  &  $12.104 $         & KIC, \cite{brown11}\tablefootmark{c}        \\
z'  &  $11.919 $         & KIC, \cite{brown11}\tablefootmark{c}        \\
Kep &  $12.495 $         & Kepler, \cite{borucki10}   \\
J   &  $10.790\pm0.026$  & 2MASS, \cite{cutri03}      \\
H   &  $10.235\pm0.030$  & 2MASS, \cite{cutri03}      \\
Ks  &  $10.136\pm0.021$  & 2MASS, \cite{cutri03}      \\
W1  &  $10.032\pm0.023$  & WISE, \cite{wright10}      \\
W2  &  $10.157\pm0.020$  & WISE, \cite{wright10}      \\
W3  &  $10.049\pm0.040$  & WISE, \cite{wright10}      \\
W4  &  $9.380  $         & WISE, \cite{wright10}      \\

\hline
\hline

\end{tabular}
\tablefoot{\\
\tablefoottext{a}{We assume 1\% of error when no errors are provided by the catalogs.}\\
\tablefoottext{b}{Johnson-like filters. More details can be found in \cite{everett12}.}\\
\tablefoottext{c}{Sloan-like filters. More details can be found in \cite{brown11}.}
}
\end{table*}

\begin{table*}[h]
\setlength{\extrarowheight}{3pt}
\caption{\label{tab:coefficients} Coefficients of the second-order polynomial fit ([Fe/H]$=a_0+a_1x+a_2x^2$) to the line absorption values in section \S~\ref{sec:metallicity} }
\centering
\begin{tabular}{rccc}
\hline\hline
Group  &  $a_0$ &  $a_1$  &  $a_2$ \\ \hline
Hot ($T_{\rm eff}>4830$ K)  & $-1.60\pm0.15$ & $0.326\pm0.046$ & $-0.0142\pm0.0036$ \\    
Cold ($T_{\rm eff}<4830$ K) &$-1.968\pm0.090$ & $0.357\pm0.022$ & $-0.0152\pm0.0013$  \\

\hline
\hline

\end{tabular}
\end{table*}

 \begin{table*}
\caption{Pulsating modes observed for Kepler-91 ordered by frequency.}             
\label{tab:frequencies}      
\centering                          
\begin{tabular}{ccc|ccc|ccc}        
\hline\hline                 

$l$ &   $\nu$   &  $\delta \nu$ & $l$ &   $\nu$   &  $\delta \nu$ & $l$ &   $\nu$   &  $\delta \nu$ \\    

\hline                        
 
  1 &   73.510  &  0.035   &  0 &   96.289  &  0.016   &  1 &  119.546  &  0.021 \\
  2 &   76.835  &  0.022   &  1 &   99.843  &  0.009   &  1 &  120.198  &  0.012 \\
  0 &   78.160  &  0.031   &  1 &  101.146  &  0.028   &  1 &  121.002  &  0.013 \\
  1 &   82.271  &  0.013   &  1 &  101.345  &  0.013   &  2 &  123.468  &  0.028 \\
  1 &   82.720  &  0.023   &  1 &  101.929  &  0.007   &  0 &  124.663  &  0.024 \\
  1 &   83.115  &  0.020   &  2 &  104.557  &  0.010   &  1 &  129.008  &  0.017 \\
  2 &   85.924  &  0.014   &  0 &  105.792  &  0.012   &  1 &  129.783  &  0.027 \\
  0 &   87.156  &  0.019   &  1 &  110.459  &  0.043   &  2 &  133.215  &  0.030 \\
  1 &   91.514  &  0.012   &  1 &  110.995  &  0.035   &  0 &  134.326  &  0.046 \\
  1 &   91.913  &  0.012   &  1 &  110.855  &  0.019   &  1 &  138.483  &  0.040 \\
  1 &   92.386  &  0.017   &  1 &  111.574  &  0.011   &  1 &  139.520  &  0.023 \\
  1 &   92.958  &  0.025   &  2 &  114.018  &  0.018   &   & & \\
  2 &   95.004  &  0.020   &  0 &  115.159  &  0.011   &   & & \\

\hline                                   
\end{tabular}
\tablefoot{The first column gives the spherical degree, the second column the frequency in $\mu$Hz, and the third column the error in the determination of the frequency. See section \S~\ref{sec:clara} for details on the method to determine these values.
}
\end{table*}

\begin{table*}[h]
\setlength{\extrarowheight}{9pt}
\caption{\label{tab:REBresults}  Results for the analysis of the primary transit and the light-curve modulations and comparison with the values obtained by \cite{tenenbaum12}, in the second column.}
\centering
\begin{tabular}{l|ccc|cc}
\hline\hline

          & \multicolumn{3}{c|}{Transit fitting}  & \multicolumn{2}{c}{REB fitting} \\  
Parameter & TCE\tablefootmark{a} & e = fixed\tablefootmark{b}   & e =free\tablefootmark{b}  &  e = fixed\tablefootmark{c}   &  e = free\tablefootmark{c} \\  
\hline
$e$                           &  0.0                            &    0.0                    & $0.13_{-0.12}^{+0.12}$             & 0.0                                         & $0.066_{-0.017}^{+0.013}$  \\        
$\omega~(^{~\circ})$          &  0.0                               &  0.0                     & $37_{-125}^{+150}$                & 0.0                                         &  $316.8_{-7.4}^{+21}$ \\              
$M_p (M_{\rm Jup})$           &  N/A\tablefootmark{d}               & N/A\tablefootmark{d}    & N/A\tablefootmark{d}              & $0.84_{-0.32}^{+0.16}$                   & $0.88_{-0.33}^{+0.17}$ \\  
$a/R_*$                       &  2.64  $\pm$ 0.23\tablefootmark{e} &   $2.40_{-0.12}^{+0.12}$   & $2.37_{-0.12}^{+0.10}$             & $2.36_{-0.35}^{+0.10}$                      &  $2.45_{-0.30}^{+0.15}$  \\  
$R_p/R_*$ $(10^{-2})$         &  2.115  $\pm$ 0.072               &   $2.255_{-0.097}^{+0.031}$   & $2.200_{-0.075}^{+0.046}$        & $2.255_{-0.097}^{+0.031}$ \tablefootmark{f} &  $2.255_{-0.097}^{+0.031}$ \tablefootmark{f} \\ 
$i~(^{~\circ})$               &     71.4  $\pm$  2.5               & $68.5_{-2.0}^{+1.0}$        &  $66.6_{-1.0}^{+2.0}$               &  $75_{-15}^{+14}$                       &  $78_{-15}^{+10}$  \\        
$\phi_{\rm offset} (10^{-3})$ &  N/A                               & $1.14_{-0.79}^{+0.74}$      & $-0.5_{-2.0}^{+0.7}$                   &  $6_{-22}^{+23}$                          &  $2_{-22}^{+23}$  \\  \hline
$\chi^2_{\rm red}$            &          3.40                      & 2.86                      &    2.86                         & 5.16                                        &  3.92    \\
$BIC$                         &              637                   &       624                 &  633                          &  1460                                       &   1411  \\


\hline
\hline

\end{tabular}
\tablefoot{\\
\tablefoottext{a}{Values from the last results of the Threshold Crossing Events by Kepler team \citep[TCE, ][]{tenenbaum12}.}\\
\tablefoottext{b}{Results from the re-analysis of the primary transit with our genetic algorithm (see section \S~\ref{sec:newfit}).}\\
\tablefoottext{c}{Results from the fitting of the REB modulations (see section \S~\ref{sec:REBfitting}).}\\
\tablefoottext{d}{Parameter not derivable by this method.}\\
\tablefoottext{e}{Derived value from the equation $r=a(1-e^2)/(1+e\cos{\xi})$ and assuming that the primary transit occurs at true anomaly $\xi=\pi/2-\omega$.}\\
\tablefoottext{f}{Assumed parameters from the transit fitting.}
}

\end{table*}

\begin{table*}[h]
\caption{\label{tab:REBparameters}  Input parameters for the
  ellipsoidal and reflection modulations fitting. We list the fixed
values and the value ranges used by the genetic algorithm to compute
the final solution.}
\centering
\begin{tabular}{rcll}
\hline\hline
\multicolumn{4}{c}{Fixed parameters (input)} \\
\hline
Parameter & Value &  Explanation &Origin \\
\hline
$T_{\rm eff}$  &  $4550\pm 75$ K &  Effective temperature & Spectroscopy (\S~\ref{sec:spec})  \\
$\log{g}$  &  $2.953\pm0.007$     &      Surface gravity  & Asteroseismology (\S~\ref{sec:mauro})  \\
$[\rm{Fe/H}]$  &  $0.11\pm0.07$     &     Stellar metallicity  & Spectroscopy (\S~\ref{sec:spec})  \\
P & $6.246580\pm0.000082$ days &  Orbital period & Light-curve \citep{batalha12}\\
$\Omega$ &  $0^{\circ}$ &  Longitude of the ascending node & Assumed \\
$\lambda_{\rm eff}^{Kepler}$  & 575 nm  & Effective wavelength of the Kepler band &   \\
$M_*$  & $1.31\pm0.10 ~ M_{\odot}$  & Stellar mass  & Asteroseismology (\S~\ref{sec:IndividualModelling}) \\ 
$R_*$  & $6.30\pm0.16 ~ R_{\odot}$  & Stellar radius  &  Asteroseismology (\S~\ref{sec:IndividualModelling})\\ 
$R_p/R_*$      & $2.255^{+0.031}_{-0.097}\times 10^{-2}$ & Planet-to-star radius & Transit fitting  (\S~\ref{sec:newfit})  \\
\hline \hline
\multicolumn{4}{c}{Free parameters (output)} \\
\hline
Parameter & [Lower limit , Upper limit] & Explanation & \\
\hline

e              & $[0.0,0.5]$                                    &  Eccentricity                            &   \\
$\omega$       & $[0^{\circ},360^{\circ}]$                      &  Argument of periastron                 &   \\
$M_p$ & $[0.5M_{\rm Jup}, 6.0M_{\rm Jup}]$     &  Planet-to-star mass ratio      &   \\
$a/R_{\star}$   & $[1.5, 5.0]$   &  Semi-major axis to stellar radius ratio &   \\
i              & $[60^{\circ},90^{\circ}]$                        &  Orbital inclination                     & \\
$\Phi_{\rm offset}$ & [-0.05,0.05] & Phase offset in phase units & \\

\hline
\hline

\end{tabular}

\end{table*}

\begin{table*}[h]
\setlength{\extrarowheight}{7pt}
\caption{ Final adopted values for the host star properties, planet and orbit of the planetary system Kepler-91.    \label{tab:finalresults} }
\centering
\begin{tabular}{lccc}
\hline\hline

Parameter & Value & Units & Method\tablefootmark{a}    \\  
\hline
Star & & \\ \hline
$T_{\rm eff}$     & $4550\pm75 $   & K  &  SP (\S~\ref{sec:teff})   \\
$[Fe/H]$ & $0.11\pm0.07 $   &  &  SP (\S~\ref{sec:metallicity})   \\
$\rho_{\star}$     & $7.3\pm0.1$   & $kg/m^3$   &  AS (\S~\ref{sec:IndividualModelling})    \\
$log(g_{\star})$     & $2.953\pm 0.007$   & cgs   &  AS (\S~\ref{sec:IndividualModelling})    \\
$M_{\star}$     &  	$1.31\pm0.10 $  &  $M_{\odot}$  &  AS (\S~\ref{sec:IndividualModelling})    \\
$R_{\star}$     &  $6.30\pm 0.16$  &  $R_{\odot}$   &  AS (\S~\ref{sec:IndividualModelling})   \\  
$L_{\star}$     &  $16.8\pm1.7$  & $L_{\odot}$ & AS (\S~\ref{sec:IndividualModelling}) \\

\hline
Planet & & & \\ \hline
$R_p/R_{\star}$ & $2.255^{+0.031}_{-0.097} \times 10^{-2}$  &   & TR (\S~\ref{sec:newfit})  \\ 
$R_p$ & $1.384^{+0.011}_{-0.054}$  & $R_{\rm Jup}$ & TR (\S~\ref{sec:newfit}) \\ 
$M_p$ & $0.88^{+0.17}_{-0.33}$ & $M_{\rm Jup}$  & REB (\S~\ref{sec:REBfitting}) \\ 
$\rho_p$ & $0.33^{+0.10}_{-0.05}$ & $\rho_{\rm Jup}$ & DER (\S~\ref{sec:discussion})  \\ 

\hline
Orbit & & & \\ \hline
i   & $68.5^{+1.0}_{-1.6}$ & deg. & REB (\S~\ref{sec:REBfitting}) \\
e   & $0.066^{+0.013}_{-0.017}$ & & REB (\S~\ref{sec:REBfitting}) \\
$a/R_{\star}$& $2.45^{+0.15}_{-0.30}$ &  & REB (\S~\ref{sec:REBfitting}) \\
$a$ & $0.072^{+0.002}_{-0.007}$ & AU & DER (\S~\ref{sec:discussion})  \\
$\omega$ & $316.8^{+21}_{-7.4}$ & deg. &  REB (\S~\ref{sec:REBfitting})  \\

\hline
\hline

\end{tabular}
\tablefoot{\\
\tablefoottext{a}{Method and section: AS = asteroseismology, SP = spectroscopy, TR = transit fitting, REB = light curve modulations fitting, and DER = derived parameters from others.}
}

\end{table*}

\end{document}